\documentclass[prl, aps, preprint, floatfix]{revtex4-1}

\usepackage[colorlinks, allcolors=blue]{hyperref}
	\hypersetup{linktocpage}

\usepackage{amsmath}
\usepackage{amssymb}
\usepackage{eufrak,amsfonts}
\usepackage{mathtools}

\usepackage{graphicx}
\usepackage{subfig}

\usepackage{xcolor}

\DeclarePairedDelimiter\ket{\lvert}{\rangle}
\DeclarePairedDelimiterX\braket[2]{\langle}{\rangle}{#1 \delimsize\vert #2}

\def\lh{\mathcal{L}}

\def\R{\mathcal{R}}

\begin{document}
\title{Spin State Dynamics in a Bichromatic Microwave Field: Role of Bright and Dark States in coupling with Reservoir}
\author{Wojciech Gawlik}
\email{gawlik@uj.edu.pl}
\author{Piotr Olczykowski}
\author{Mariusz Mr\'ozek}
\author{Adam M. Wojciechowski}
\affiliation{Institute  of  Physics,  Jagiellonian  University, {\L}ojasiewicza  11,  30-348  Krak\'ow,  Poland}

\begin{abstract}
Two nearly degenerate fields acting on an open spin system enable observation of composite magnetic resonances with nontrivial intensity dependence, like immunity to power broadening. Interaction of a spin system with bichromatic microwave field specifies linear combinations of populations interpreted in terms of coupled (bright) and uncoupled (dark) states, provides evidence of openness of the system, and enables controlling their coupling with reservoir of probability.

\end{abstract}

\maketitle

Driving spin systems with specific time-dependent fields enables engineering of spin ensembles, observation of many interesting phenomena, and important applications. A fundamental condition for such developments is a sound understanding of their relaxation and decoherence and the ability to control interactions with other fields or systems.  Significant progress in such control has been achieved by coherent dark state formation  \cite{gray1978,arimondo1996,santori2006,xu2008,togan2011,hansom2014,yale2013,rogers2014,pingault2014,jamonneau2016}, and engineering of quantum states and their environments \cite{smeltzer2009,englund2010,de2012,belthangady2013,childress2013,northup2014,rohr2014,golter2014,lee2017,gu2017,choi2017,kucsko2018,astner2018,bauch2018,barry2020}.

Below, we present results of our theoretical and experimental study of bichromatic driving an open two-level spin system where application of the hole burning \cite{kehayias2014,putz2017,soltamov2019} and coherent population oscillations (CPO) \cite{mrozek2016,baklanov1972,sargent1978,laupretre2012,ella2019} methodology enables observation of composite, i.e. nested, multicomponent resonances. When analyzing shapes of the composite resonances versus the microwave (MW) power, we observe a peculiar effect that one component of the structure exhibits regular power broadening, whereas the other one becomes broadening-free, i.e. stabilized. We explain the observed stabilization by analyzing the dynamics of combinations of spin state populations and demonstrating specific power-dependent linear combinations decoupled from each other in the time-independent regime of the evolution generator. For MW power close to zero, they reproduce the differences between initial populations and stationary solution while for strong MWs they converge to one combination with its dynamics proportional to MW power (bright state) and another one with dynamics independent on MWs (dark state). Our approach is similar to that applied in CPT, with an important difference that here we consider combinations of populations, rather than wave functions or coherences. The analyzed resonances provide a sensitive indication of the openness of the system and can be used for characterization of spin dynamics in various paramagnetic samples and control their interaction with external fields.

\begin{figure}[tb]
	\centering
	\includegraphics[width=0.8\linewidth]{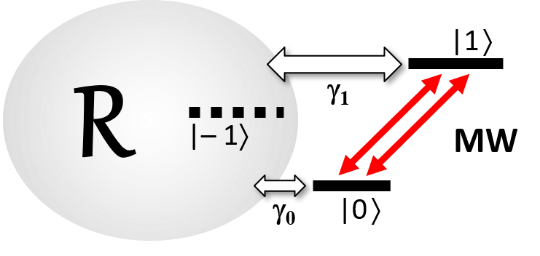}
	\caption{The model system of an open \((S=1)\) spin system driven by two MW fields nearly resonant with  transition, \( \ket{0} - \ket{1}\), and coupled  to  reservoir \(\R\) with rates \(\gamma_1=\gamma(1+\epsilon)\), \(\gamma_0=\gamma(1-\epsilon)\). In appropriately strong magnetic field state \(\ket{-1}\) (marked with a broken line), non-resonant with the MWs, is treated as a part of \(\R\) and the analysis reduces to a two-level system. }

	\label{fig:model}
\end{figure}

\begin{figure}[tb]
	\centering
	\includegraphics[width=0.9\linewidth]{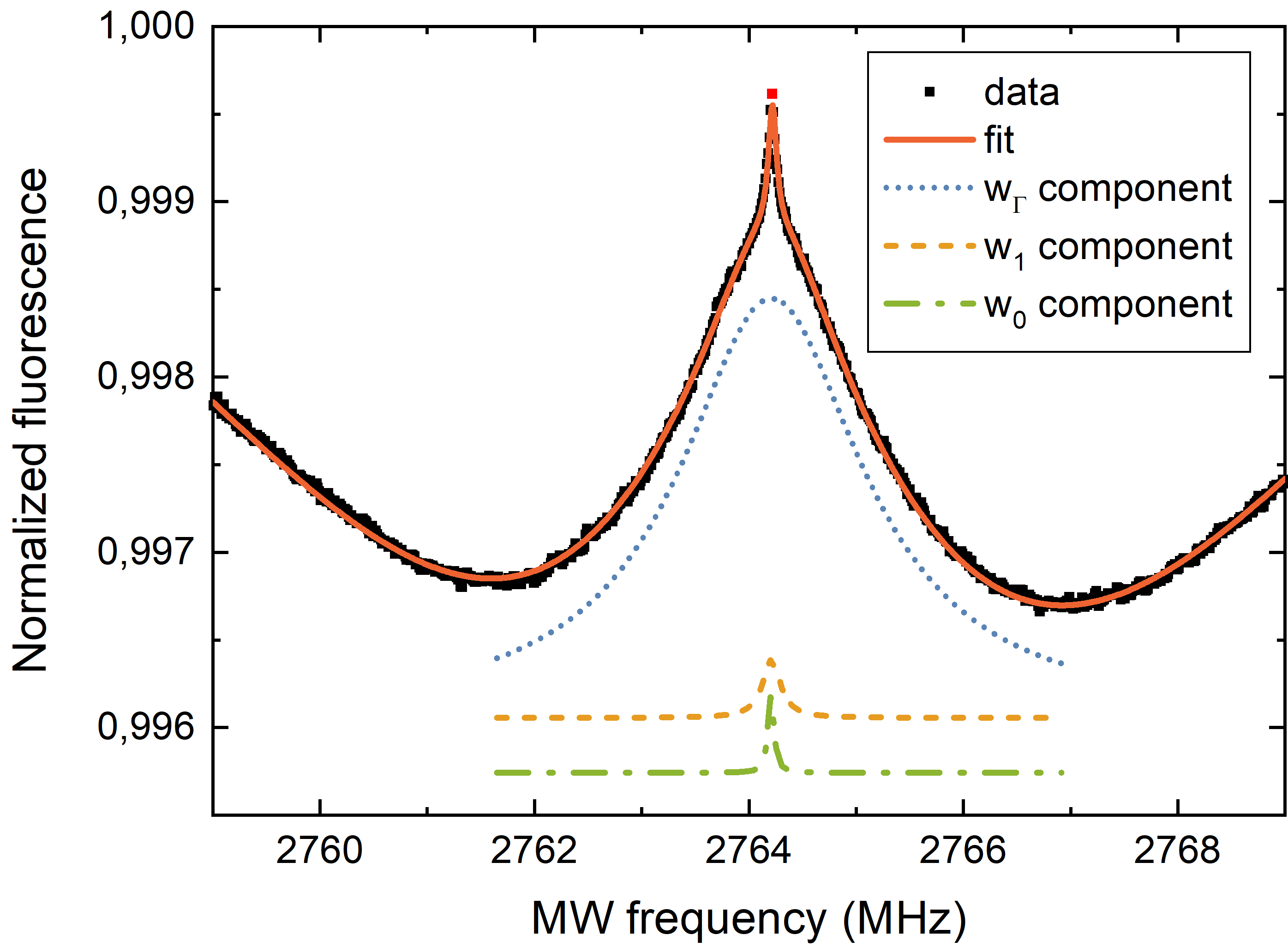}
	\caption{ Experimental data (for \(\omega_2=\omega_0\)) fitted with a composite triple-Lorenz curve on a Gaussian background.}
	\label{fig:expdata}
\end{figure}

Our modelling is verified experimentally with an ensemble of nitrogen vacancy (NV) color centers in diamond crystal excited by a green light and driven by two microwave fields of comparable strengths and nearly resonant frequencies. Each NV center is a multilevel spin system interacting with other NV centers, crystal impurities, etc. As described in Refs. \cite{kehayias2014,mrozek2016}, such situation enables observation of holes burnt in inhomogeneously broadened optically detected magnetic resonance (ODMR) profile. Two different situations may be realized: (i) one of the MW fields is tuned to one transition, e.g. between spin states \(\ket{m_S=0}\)   and \(\ket{m_S=-1}\)  while the second MW is close to resonance with another one, e.g. \(\ket{m_S=0} \leftrightarrow \ket{m_S=+1}\) \cite{kehayias2014}, and (ii) both MWs are close to resonance with one transition between \(\ket{m_S=0}\) and either of spin states of the ground state \cite{mrozek2016}. In this Letter, we focus on the second case when CPO with frequency \(\delta=\omega_1-\omega_2\) take place between the coupled spin states and result in a composite resonance shape (Fig.~\ref{fig:expdata}) which in the lowest order is composed of three resonance contributions associated with specific decays of populations of the two spin states and their coherence. In magnetic fields of a few mT, aligned with particular NV axis, the  \(\ket{m_S=0} \leftrightarrow \ket{m_S=\pm 1}\) transitions of the NV center are well resolved, thus we are left with an effective two-level subsystem \(\ket{m_S=0}\), \(\ket{m_S=+ 1}\), denoted as \(\ket{0}\) and \(\ket{1}\), respectively, driven by two quasi-resonant MW fields (frequencies \(\omega_1,\,\omega_2\)) and coupled through relaxation rates \(\gamma_0\), \(\gamma_1\) with a reservoir of probability \(\R\), which consists of all other undetected NV energy levels,  including the non-resonant \(\ket{-1}\) one (Fig. \ref{fig:model}). 

Since the reservoir is just the complement of examined spin state populations, it can be included in the evolution equations of a two-level system as self-coupling (relaxation to equilibrium) of populations and coherence via master equation: \( \dot \rho_{kk}=-\gamma_k(\rho_{kk}-\rho_{kk}^0) \) and \( \dot\rho_{jk}=-\Gamma\rho_{jk},\) where \(\rho\) is a density matrix (\(j,k = 0,1\) and \(j\neq k\)) and the \(\gamma_k\) rates are expressed with the help of asymmetry parameter \(\epsilon\), via \(\gamma_k=\gamma(1-(-1)^k\epsilon)\) with \(\gamma\) being average of population relaxation rates. If both populations relax with the same rate, i.e.  \(\gamma_0=\gamma_1\) or \(\epsilon=0\), the whole probability is conserved and such system is referred to as a closed two-level system. On the other hand, if \(\epsilon\neq0\), the whole probability is unequally flowing in and out of the system so we  refer to it as an open two-level system. 

Adiabatic elimination of coherence \(\rho_{jk}\) (for \(\delta\ll \Gamma\)) reduces the analysis to studying the dynamics of two populations, \(n_k:=\rho_{kk}\), oscillating with frequency \(\delta\) and the amplitude depending on MW strength and detuning. Consequently, the population dynamics reflects interplay of exponential decay and CPO. We represent it as a reduced master equation for populations only, \(n_k \in [0,1]\):

\begin{equation}\begin{split} \label{eq:dalet}
\dfrac{d}{d(\gamma t)} n_1= - (1+\epsilon)(n_1-n_1^0) - S\big(1 + \cos \delta t \big) (n_1-n_0),\\
\dfrac{d}{d(\gamma t)} n_0= - (1-\epsilon)(n_0-n_0^0) +  S\big(1 + \cos \delta t \big)  (n_1-n_0),
\end{split}\end{equation}
where we take \(\delta\ll\Gamma\), assume both MWs equally strong and characterized by Rabi frequency \(\Omega\), define \(S=\frac{\Omega^2}{\gamma\Gamma}\cdot\left( L_\Gamma(\omega_1-\omega_0)+ L_\Gamma(\omega_2-\omega_0)\right)\) as the saturation parameter, and use a normalized Lorentz function \(L_a(x)=\frac{a^2}{x^2+a^2}\). Details can be found in \cite{suplement}. We represent the fluorescence signal as a time averaged linear combination of populations. Consequently, the composite resonance strongly depends on the dimensionality of the solution of Eq.~\ref{eq:dalet} which limits the number of components contributing to the resonance. For \( \epsilon=0 \), one of the equations becomes redundant since the total probability is conserved and only the population difference evolves. Thus, the dynamics of the system is constrained to a one-dimensional (1D) subspace and the resonance has a form of a single hole power-broadened for all values of \( S \). On the contrary, for \( \epsilon \neq 0 \), the sum of populations is not constant in time, the system is open, i.e. its dynamics is two-dimensional (2D).

Below, we show that the system's dynamics can be described in terms of two independent combinations of populations, strongly and weakly self-coupled through the MW driving. Eq. \ref{eq:dalet} describing nonunitary evolution of populations is a first order linear differential equation that, to some extent, resembles a standard quantum state evolution in the interaction picture \cite{arimondo1996,gray1978}, yet with two important differences: (i) the evolving states are not complex superpositions of basis vectors but real superpositions of populations, and (ii) the evolution operator generated by the integration of the right-hand side of Eq. \ref{eq:dalet} is not a unitary one.
We identify the widths of the resonance components as the signatures of two significantly different lifetimes of distinct states and the discussed phenomenon as a nonunitary, classical analog to the CPT effect. Eq. \ref{eq:dalet} constitute thus a classical, rather than quantum, form of a Liouville equation. The Liouvillian may be decomposed into a sum of time-independent \( \mathcal{L}_0 \) and time-dependent \( \mathcal{L}_1(t) \) parts: 

\begin{equation}\label{eq:he} 
\dot n   = - (\mathcal{L}_0+\mathcal{L}_1(t))\cdot n + n^0,
\end{equation} 
with:

\[\begin{split}
n & = \begin{pmatrix}n_1 \\n_0\end{pmatrix}, \quad
\mathcal{L}_0=\begin{pmatrix}1+S+\epsilon & -S \\-S & 1+S-\epsilon\end{pmatrix},\\
n^0&= \begin{pmatrix}(1+\epsilon)n_1^0 \\(1-\epsilon)n_0^0\end{pmatrix},
\quad \mathcal{L}_1(t)=S \cos \delta t \cdot \begin{pmatrix}1 & -1 \\-1 & 1\end{pmatrix}.
\end{split}\]

Operators \(\mathcal{L}_0\) and \(\mathcal{L}_1\) play significantly different roles in the dynamics of the system. In our description \(\mathcal{L}_0\) is responsible for dressing the initial populations, whereas \(\mathcal{L}_1\) provides a time dependent coupling between them. The formal solution to the time-independent dressing part of Eq. \ref{eq:he} can be written as:

\begin{equation}\label{eq:waw1}
n(t)=e^{-\mathcal{L}_0\cdot\gamma t}(n(0)-\bar n)+ \bar n,
\end{equation}
where \(\bar n \colon= \mathcal{L}_0^{-1}n^0\) denotes the stationary solution. Diagonalization of \(\mathcal{L}_0\) yields eigenvalues: \(\lambda_k=1+S-(-1)^k\sqrt{S^2+\epsilon^2}\). The  difference between populations and stationary solution is represented in the \(\mathcal{L}_0-\) eigenbasis:
 
\begin{equation}\label{eq:waw}
 \eta(t) \colon = \begin{pmatrix}
e^{-\lambda_1 \cdot\gamma t}\eta_1(0) \\e^{-\lambda_0\cdot\gamma t}\eta_0(0)
\end{pmatrix},
\end{equation}
where \(\eta_k(0)\) are determined by initial conditions, \(\eta_1(t)=\cos \theta (n_1-\bar n_1^0) - \sin \theta (n_0-\bar n_0^0 )\) and \(\eta_0=\sin \theta (n_1-\bar n_1^0)+\cos \theta (n_0-\bar n_0^0)\) with \(\theta\) being the mixing angle.
     
Eqs. \ref{eq:waw1} and \ref{eq:waw} suggest  interpretation of combinations  \(\eta_1\) and \(\eta_0\) as two states with coefficients \(\gamma\cdot \lambda_k = \gamma\cdot (1+S-(-1)^k\sqrt{S^2+\epsilon^2})\) being their relaxation rates. We discover here an analogy with CPT and the familiar interpretation of the bright (coupled) and dark (uncoupled) states \cite{arimondo1996,gray1978}. Note, that \(\eta_{k}\) are combinations of populations, rather than coherent superpositions of wavefunctions as in a standard CPT formalism. For strong MWs (\(S\gg1\)), \(\lambda_1 \simeq 1+2S\) and \(\lambda_0\simeq 1\), hence, we interpret these states as strongly and weakly coupled to the reservoir and, respectively, short and long living or power-broadened and power-stabilized.

\begin{figure}[t!]
	\centering
	\includegraphics[width=0.9\linewidth]{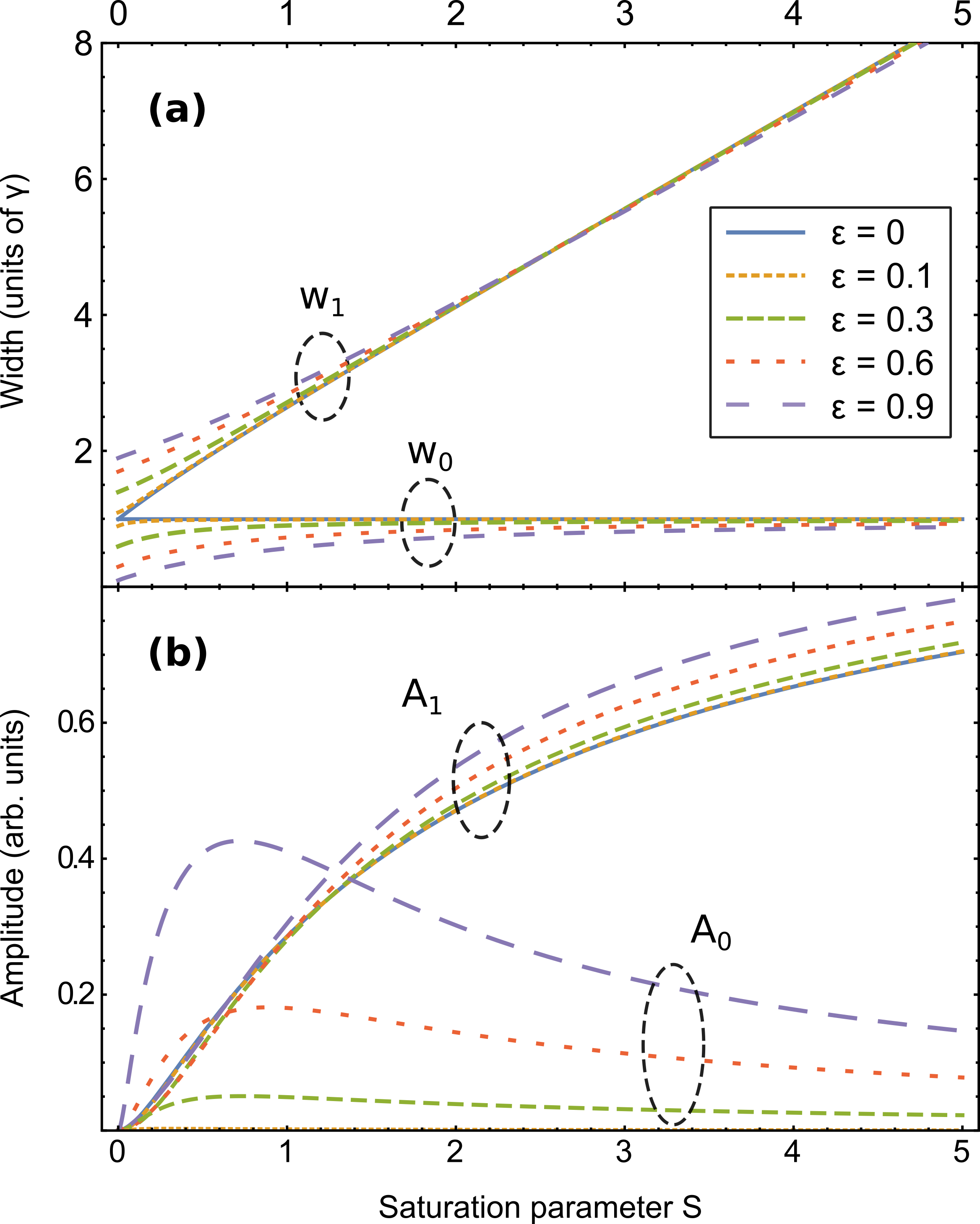}
	\caption{Amplitudes \(A_0\), \(A_1\) (a) and widths \(w_0\), \(w_1\) (b) of the resonance components calculated within first-harmonic approximation for various values of assymmetry parameter \(\epsilon\).}  
	\label{fig:ampwidths}
\end{figure}

In the absence of the second MW field, the negative self-coupling is responsible for exponential decay, thus the difference between coupling constants may explain multiexponential decays observed in  \cite{choi2017,kucsko2018,mrozek2015}. However, if MW field exhibits a time modulation, like in our bichromatic case, emerging CPO confronts exponential decay and the two coupling constants are reflected in the multicomponent composite resonance.

To provide a measurable prediction on the widths of the components of the resonance we exploit the solution of full, time dependent Eq. \ref{eq:he}. For this sake, we expand the populations into a Fourier series and interpret its components as amplitudes of the harmonics of \(\delta\). For sufficiently low power, \(\Omega\ll\Gamma\), we focus on the first harmonic and get for the time-averaged  difference \(\langle\Delta n \rangle = \langle n_1-n_0\rangle\) \cite{suplement}:

\begin{equation}\label{eq:zain}
 I \sim   \frac{\Delta n^0}{1+\frac{2S}{1-\epsilon^2}}\cdot \left(1-A_0 \,L_{w_0}(\delta)-A_1 \,L_{w_1}(\delta)\right),
\end{equation}
where \(\Delta n^0\) is the initial population difference, and \(A_{k}\) and \(w_{k}\) are functions of \(S\) and \(\epsilon \) only \cite{suplement}. The first term in Eq. \ref{eq:zain} yields a background which for \(\Gamma\gg\gamma\) appears as flat but for \(\Gamma\gtrsim \gamma\) represents a hole of width of the order \( (T_2^*)^{-1} \)  (ODMR) burnt by two independent MWs \cite{kehayias2014}. The next terms in Eq. \ref{eq:zain} correspond to narrow components of the composite resonance caused by CPO, represented by two Lorentzians with the amplitudes and widths exhibiting different dependencies on \(S\): \(A_1\) and \(w_1\) rise monotonically, whereas \(A_0\) first rises but beyond \(S\simeq 1\) falls down to \(0\) while \(w_0\) slowly saturates at \(\gamma\) when \(S\) approaches infinity (Fig. \ref{fig:ampwidths}).

For a closed system (\(\epsilon=0\)), the structure of the resonance simplifies to a single power-broadened resonance with its width nearly linearly dependent on \(S\) (in Fig. \ref{fig:ampwidths}a), but for an open system the composite resonance consists of two contributions with different widths and amplitudes. For very weak MW power (\(S\ll 1\)), both contributions are power broadened with equal rates but for \(S\geq 1\) one component remains strongly power broadened, while another one becomes independent on \(S\), i.e.  stabilizes. For strong MWs \((S \geq 1)\), \(w_{k}\) acquire simple forms: 
\begin{equation}
    w_0\simeq \gamma \cdot \lambda_0 ,\quad w_1\simeq \gamma \cdot \sqrt{\lambda_1^2-2S^2},
\end{equation}
which in the limit \(S\gg 1\) yields: \( w_0 \simeq\gamma\)  and \(w_1 \simeq \sqrt{2} \gamma S +\frac{\gamma(4-\epsilon^2)}{2\sqrt{2}} \) have been neglected) clearly demonstrating stabilization of \(w_0\) at the mean value \(\gamma\). This effect is a direct consequence of the existence of the dark state \(\eta_0\) and constitutes the main finding of this work. 

We have verified our theory with a CW ODMR setup and laser-excited (\(532\)~nm light) NV diamond ensemble ([NV]\(\sim\)10~ppm) in a magnetic field of \(4\)~mT \cite{mrozek2016,kehayias2014}. A superposition of three Lorentzians was fitted to the composite resonance recorded with two MWs tuned to the same \(\ket{m_S=0}\leftrightarrow\ket{m_s=1}\) transition. Figure~\ref{fig:expdata} demonstrates very good agreement of the calculations with the observations presented in Ref. \cite{mrozek2016}.   In particular, the ODMR resonance for bichromatic driving with two MWs of nearly the same frequencies exhibits a hole (note that the ODMR resonance is a dip, hence "holes" appear as peaks)  composed of three contributions, reproduced in Eq. \ref{eq:zain}, with amplitudes \(A_0\), \(A_1\) and widths \(w_0\), \(w_1\)  and \(w_\Gamma\), which we associate with the population relaxation rates \(\gamma_0\), \(\gamma_1\) and decoherence \(\Gamma\), respectively. 

\begin{figure}
    \includegraphics[width=0.9\linewidth]{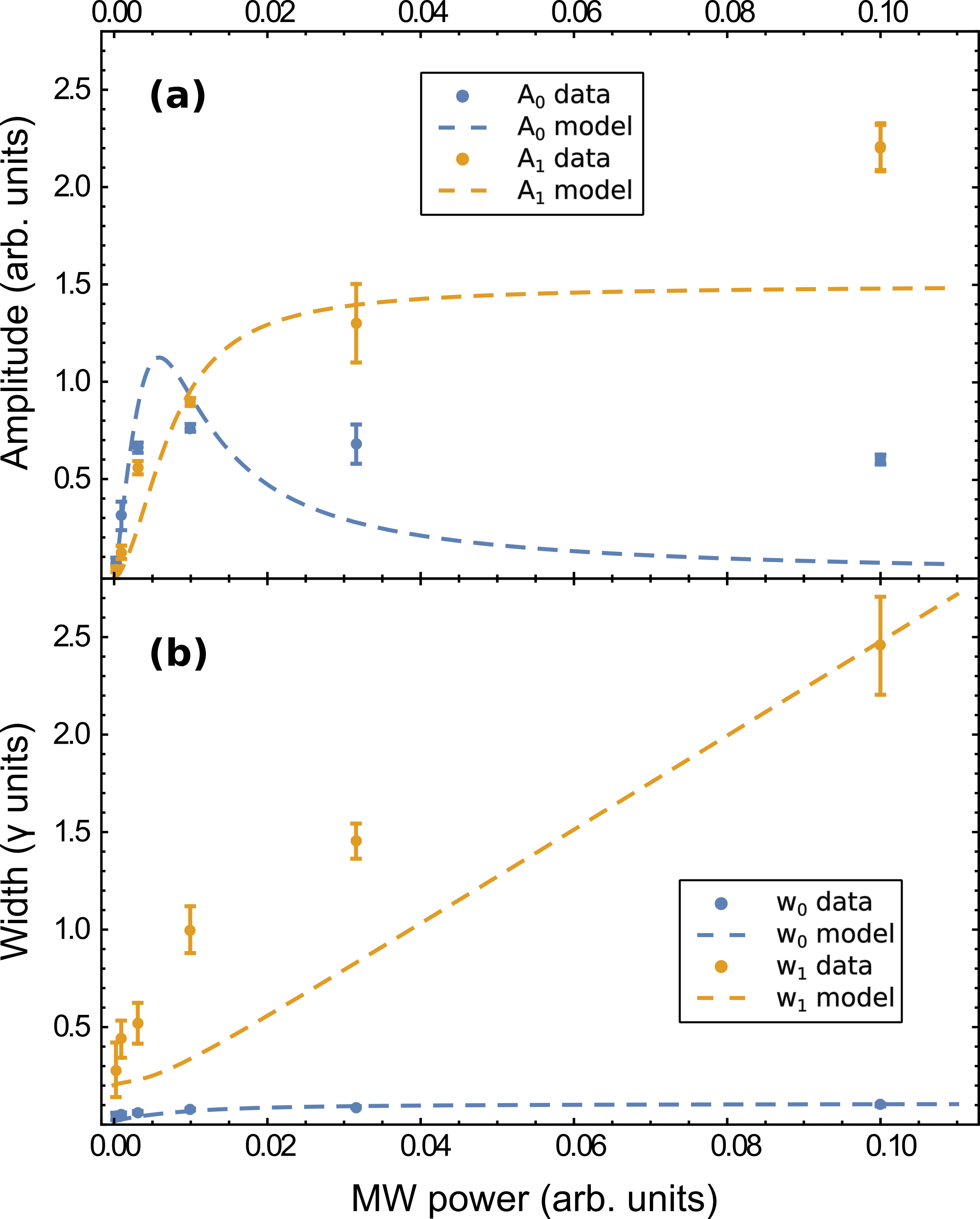}

	\caption{Measured properties of the composite resonance vs. MW power recorded with \(\omega_2=\omega_0\) (points and error bars) and theoretical predictions (solid lines) for \(\gamma = 0.11\) and \(\epsilon=0.85\): (a) amplitudes \(A_0,\,A_1\); (b) widths \(w_0,\,w_1\).}
	\label{fig:plots}
\end{figure}

Figure \ref{fig:plots} presents amplitudes \(A_0\), \(A_1\) (Fig. \ref{fig:plots}a) and widths \(w_0\), \(w_1\) (Fig. \ref{fig:plots}b) of the resonance components measured as a function of MW power. In agreement with our theory we observe strong broadening of \(w_1\) and almost no dependence of \(w_0\) on \(S\)  which confirms the predicted \emph{light-induced stabilization} of population of one of the superposition states. There is also a qualitative agreement of the amplitude dependences  \(A_0(S)\) and \(A_1(S)\). While the predicted stabilization of \(w_0(S)\) is well evidenced in Fig.\ref{fig:plots}, quantitative modelling of \(w_1(S)\) and \(A_k(S)\) is far from being perfect. This is caused by the departure from our simplifying assumption \(\Gamma\gg\gamma\).  In fact, in the experiment \(\Gamma\simeq 10\cdot \gamma\) which partly violates our adiabatic approximation and results in a non-negligible overlap of the regular hole (first term of \ref{eq:zain}) with the composite resonance. Moreover, for \(S\gtrsim1\), higher harmonics of \(\delta\) become relevant. Consequently, modelling of \(w_k\) and \(A_k\) is not very accurate. Still, when \(\gamma_0<\gamma_1\), as in our case, the resulting error is smaller when modeling \(w_0\) than \(w_1\) and \(A_k\) which is seen by a very good agreement of \(w_0(S)\) with the theoretical prediction. In fact, a more accurate analysis with accounting for higher harmonics (to be published elsewhere) also confirms the above described stabilization effect.   

\begin{figure}
	\centering
	\includegraphics[width=0.9\linewidth]{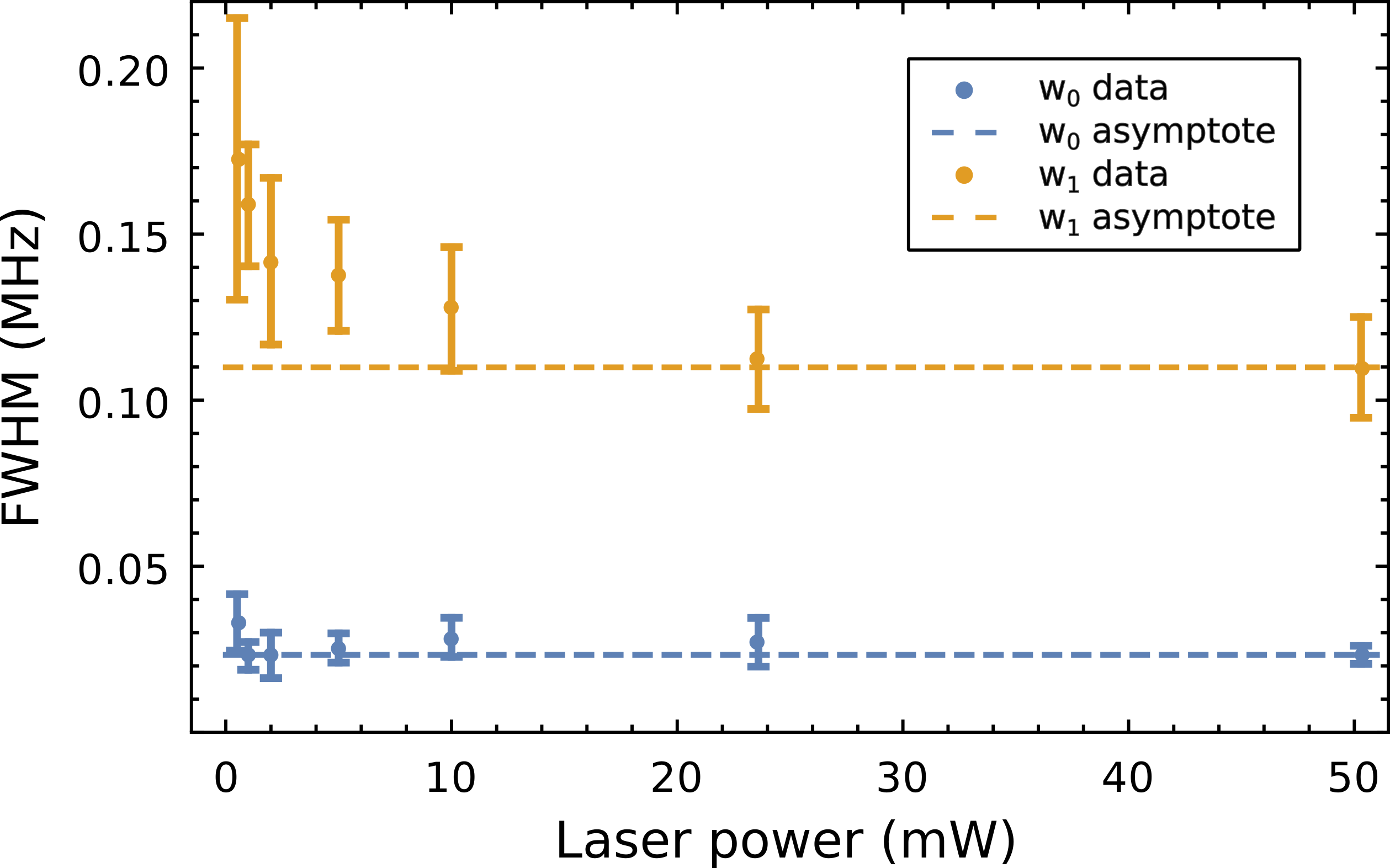}
	\caption{ Light-narrowing effect the of resonance widths for a fixed MW power. }
	\label{fig:narrowing}
\end{figure}

Another important assumption of our model is the reduction of the role of optical pumping to the mere establishment of the initial spin polarization \(\Delta n^0\). To get more insight on that role, we studied experimentally the effect of light power \(P_{\mathrm{light}}\) on the measured spin dynamics. Figure \ref{fig:narrowing} shows the results of measurements of widths \(w_0\) and \(w_1\) as a function of CW light power for constant MW power. Similarly to Fig. \ref{fig:plots}, in the applied range of light intensities we did not observe any significant change of \(w_\Gamma\) and \(w_0\), which demonstrates stabilization of populations against light perturbation \cite{dreau2011}. On the other hand, a twofold narrowing of \(w_1\) is clearly visible. With more intense light both resonance widths become saturated and do not depend on \(P_{\mathrm{light}}\). Similarly as in \cite{jensen2013}, we ascribe this narrowing to the effect of optical pumping and intersystem crossing which involves at least five energy levels. In contrast to previous studies, the narrowing reported here addresses a narrow (well below \( (T_2^*)^{-1} \) linewidth) spectral feature and occurs for much lower light intensities (\(P_\mathrm{light}<20\)mW).

In summary, we have presented the theoretical and experimental analysis of an intriguing phenomenon of the field-induced stabilization of composite resonances created by bichromatic excitation.  Specifically, we have found that the two-field methodology enables addressing of individual spin states and studies of their interaction with reservoir, which is impossible with standard cw ODMR experiment where the resonances are jointly  affected by both relaxation rates. The very existence of a multicomponent resonance revealed by bichromatic spectroscopy implies that the system under consideration is an open one. The method serves thus as a sensitive indicator of the system's openness. The states identified by different relaxation times result from the competition between relaxation and MW driving. For the rising MW power, the two dressed states tend to the sum and difference of the initial populations: the dynamics of the population difference is governed by the MW power while the population sum becomes power-independent and an initially open system effectively closes for strong MWs (since \(\det \lh_1=0\) for all \(S\)). This fact is manifested experimentally as a decrease of \(A_0\) with rising MWs. The described effect of the width stabilization of one component of the composite resonance and power broadening of another one at high MW power has a similar origin to that seen in CPT: in our case, the role of the strongly (bright) and weakly coupled (dark) states is played by population superpositions \(\eta_1\) and \(\eta_0\), respectively. 

The reported theoretical analysis and measurements provide an insight into the structure of the population dynamics of open systems. Although our experiment dealt with a special case of NV diamond, the analysis and discovered phenomena are general and may be useful for characterization of spin dynamics of various paramagnetic samples and control of their interaction with external fields. The described reduction of power broadening and resulting narrow resonance width should be helpful for quantitative characterization of relaxation mechanism and precision spectroscopy of the studied spin systems. For such applications, the investigated samples should possess possibly different population relaxation rates and much smaller than the overall dephasing rate: \(\gamma_0<\gamma_1\ll\Gamma\).  

\begin{acknowledgments}
Authors acknowledge financial support by the National Science Centre, Poland (grant number 2016/21/B/ST7/01430) and~Foundation for Polish Science (grant number POIR~04.04.00-00-1644/18).
\end{acknowledgments}


\begin{thebibliography}{37}%
\makeatletter
\providecommand \@ifxundefined [1]{%
 \@ifx{#1\undefined}
}%
\providecommand \@ifnum [1]{%
 \ifnum #1\expandafter \@firstoftwo
 \else \expandafter \@secondoftwo
 \fi
}%
\providecommand \@ifx [1]{%
 \ifx #1\expandafter \@firstoftwo
 \else \expandafter \@secondoftwo
 \fi
}%
\providecommand \natexlab [1]{#1}%
\providecommand \enquote  [1]{``#1''}%
\providecommand \bibnamefont  [1]{#1}%
\providecommand \bibfnamefont [1]{#1}%
\providecommand \citenamefont [1]{#1}%
\providecommand \href@noop [0]{\@secondoftwo}%
\providecommand \href [0]{\begingroup \@sanitize@url \@href}%
\providecommand \@href[1]{\@@startlink{#1}\@@href}%
\providecommand \@@href[1]{\endgroup#1\@@endlink}%
\providecommand \@sanitize@url [0]{\catcode `\\12\catcode `\$12\catcode
  `\&12\catcode `\#12\catcode `\^12\catcode `\_12\catcode `\%12\relax}%
\providecommand \@@startlink[1]{}%
\providecommand \@@endlink[0]{}%
\providecommand \url  [0]{\begingroup\@sanitize@url \@url }%
\providecommand \@url [1]{\endgroup\@href {#1}{\urlprefix }}%
\providecommand \urlprefix  [0]{URL }%
\providecommand \Eprint [0]{\href }%
\providecommand \doibase [0]{http://dx.doi.org/}%
\providecommand \selectlanguage [0]{\@gobble}%
\providecommand \bibinfo  [0]{\@secondoftwo}%
\providecommand \bibfield  [0]{\@secondoftwo}%
\providecommand \translation [1]{[#1]}%
\providecommand \BibitemOpen [0]{}%
\providecommand \bibitemStop [0]{}%
\providecommand \bibitemNoStop [0]{.\EOS\space}%
\providecommand \EOS [0]{\spacefactor3000\relax}%
\providecommand \BibitemShut  [1]{\csname bibitem#1\endcsname}%
\let\auto@bib@innerbib\@empty
\bibitem [{\citenamefont {Gray}\ \emph {et~al.}(1978)\citenamefont {Gray},
  \citenamefont {Whitley},\ and\ \citenamefont {Stroud}}]{gray1978}%
  \BibitemOpen
  \bibfield  {author} {\bibinfo {author} {\bibfnamefont {H.~R.}\ \bibnamefont
  {Gray}}, \bibinfo {author} {\bibfnamefont {R.~M.}\ \bibnamefont {Whitley}}, \
  and\ \bibinfo {author} {\bibfnamefont {C.~R.}\ \bibnamefont {Stroud}},\
  }\href@noop {} {\bibfield  {journal} {\bibinfo  {journal} {Optics letters}\
  }\textbf {\bibinfo {volume} {3}},\ \bibinfo {pages} {218} (\bibinfo {year}
  {1978})}\BibitemShut {NoStop}%
\bibitem [{\citenamefont {Arimondo}(1996)}]{arimondo1996}%
  \BibitemOpen
  \bibfield  {author} {\bibinfo {author} {\bibfnamefont {E.}~\bibnamefont
  {Arimondo}},\ }in\ \href@noop {} {\emph {\bibinfo {booktitle} {Progress in
  Optics}}},\ Vol.~\bibinfo {volume} {35},\ \bibinfo {editor} {edited by\
  \bibinfo {editor} {\bibfnamefont {E.}~\bibnamefont {Wolf}}}\ (\bibinfo
  {publisher} {Elsevier},\ \bibinfo {year} {1996})\ pp.\ \bibinfo {pages}
  {257--354}\BibitemShut {NoStop}%
\bibitem [{\citenamefont {Santori}\ \emph {et~al.}(2006)\citenamefont
  {Santori}, \citenamefont {Tamarat}, \citenamefont {Neumann}, \citenamefont
  {Wrachtrup}, \citenamefont {Fattal}, \citenamefont {Beausoleil},
  \citenamefont {Rabeau}, \citenamefont {Olivero}, \citenamefont {Greentree},
  \citenamefont {Prawer} \emph {et~al.}}]{santori2006}%
  \BibitemOpen
  \bibfield  {author} {\bibinfo {author} {\bibfnamefont {C.}~\bibnamefont
  {Santori}}, \bibinfo {author} {\bibfnamefont {P.}~\bibnamefont {Tamarat}},
  \bibinfo {author} {\bibfnamefont {P.}~\bibnamefont {Neumann}}, \bibinfo
  {author} {\bibfnamefont {J.}~\bibnamefont {Wrachtrup}}, \bibinfo {author}
  {\bibfnamefont {D.}~\bibnamefont {Fattal}}, \bibinfo {author} {\bibfnamefont
  {R.~G.}\ \bibnamefont {Beausoleil}}, \bibinfo {author} {\bibfnamefont
  {J.}~\bibnamefont {Rabeau}}, \bibinfo {author} {\bibfnamefont
  {P.}~\bibnamefont {Olivero}}, \bibinfo {author} {\bibfnamefont {A.~D.}\
  \bibnamefont {Greentree}}, \bibinfo {author} {\bibfnamefont {S.}~\bibnamefont
  {Prawer}},  \emph {et~al.},\ }\href@noop {} {\bibfield  {journal} {\bibinfo
  {journal} {Physical review letters}\ }\textbf {\bibinfo {volume} {97}},\
  \bibinfo {pages} {247401} (\bibinfo {year} {2006})}\BibitemShut {NoStop}%
\bibitem [{\citenamefont {Xu}\ \emph {et~al.}(2008)\citenamefont {Xu},
  \citenamefont {Sun}, \citenamefont {Berman}, \citenamefont {Steel},
  \citenamefont {Bracker}, \citenamefont {Gammon},\ and\ \citenamefont
  {Sham}}]{xu2008}%
  \BibitemOpen
  \bibfield  {author} {\bibinfo {author} {\bibfnamefont {X.}~\bibnamefont
  {Xu}}, \bibinfo {author} {\bibfnamefont {B.}~\bibnamefont {Sun}}, \bibinfo
  {author} {\bibfnamefont {P.~R.}\ \bibnamefont {Berman}}, \bibinfo {author}
  {\bibfnamefont {D.~G.}\ \bibnamefont {Steel}}, \bibinfo {author}
  {\bibfnamefont {A.~S.}\ \bibnamefont {Bracker}}, \bibinfo {author}
  {\bibfnamefont {D.}~\bibnamefont {Gammon}}, \ and\ \bibinfo {author}
  {\bibfnamefont {L.}~\bibnamefont {Sham}},\ }\href@noop {} {\bibfield
  {journal} {\bibinfo  {journal} {Nature Physics}\ }\textbf {\bibinfo {volume}
  {4}},\ \bibinfo {pages} {692} (\bibinfo {year} {2008})}\BibitemShut {NoStop}%
\bibitem [{\citenamefont {Togan}\ \emph {et~al.}(2011)\citenamefont {Togan},
  \citenamefont {Chu}, \citenamefont {Imamoglu},\ and\ \citenamefont
  {Lukin}}]{togan2011}%
  \BibitemOpen
  \bibfield  {author} {\bibinfo {author} {\bibfnamefont {E.}~\bibnamefont
  {Togan}}, \bibinfo {author} {\bibfnamefont {Y.}~\bibnamefont {Chu}}, \bibinfo
  {author} {\bibfnamefont {A.}~\bibnamefont {Imamoglu}}, \ and\ \bibinfo
  {author} {\bibfnamefont {M.}~\bibnamefont {Lukin}},\ }\href@noop {}
  {\bibfield  {journal} {\bibinfo  {journal} {Nature}\ }\textbf {\bibinfo
  {volume} {478}},\ \bibinfo {pages} {497} (\bibinfo {year}
  {2011})}\BibitemShut {NoStop}%
\bibitem [{\citenamefont {Hansom}\ \emph {et~al.}(2014)\citenamefont {Hansom},
  \citenamefont {Schulte}, \citenamefont {Le~Gall}, \citenamefont {Matthiesen},
  \citenamefont {Clarke}, \citenamefont {Hugues}, \citenamefont {Taylor},\ and\
  \citenamefont {Atat{\"u}re}}]{hansom2014}%
  \BibitemOpen
  \bibfield  {author} {\bibinfo {author} {\bibfnamefont {J.}~\bibnamefont
  {Hansom}}, \bibinfo {author} {\bibfnamefont {C.~H.}\ \bibnamefont {Schulte}},
  \bibinfo {author} {\bibfnamefont {C.}~\bibnamefont {Le~Gall}}, \bibinfo
  {author} {\bibfnamefont {C.}~\bibnamefont {Matthiesen}}, \bibinfo {author}
  {\bibfnamefont {E.}~\bibnamefont {Clarke}}, \bibinfo {author} {\bibfnamefont
  {M.}~\bibnamefont {Hugues}}, \bibinfo {author} {\bibfnamefont {J.~M.}\
  \bibnamefont {Taylor}}, \ and\ \bibinfo {author} {\bibfnamefont
  {M.}~\bibnamefont {Atat{\"u}re}},\ }\href@noop {} {\bibfield  {journal}
  {\bibinfo  {journal} {Nature Physics}\ }\textbf {\bibinfo {volume} {10}},\
  \bibinfo {pages} {725} (\bibinfo {year} {2014})}\BibitemShut {NoStop}%
\bibitem [{\citenamefont {Yale}\ \emph {et~al.}(2013)\citenamefont {Yale},
  \citenamefont {Buckley}, \citenamefont {Christle}, \citenamefont {Burkard},
  \citenamefont {Heremans}, \citenamefont {Bassett},\ and\ \citenamefont
  {Awschalom}}]{yale2013}%
  \BibitemOpen
  \bibfield  {author} {\bibinfo {author} {\bibfnamefont {C.~G.}\ \bibnamefont
  {Yale}}, \bibinfo {author} {\bibfnamefont {B.~B.}\ \bibnamefont {Buckley}},
  \bibinfo {author} {\bibfnamefont {D.~J.}\ \bibnamefont {Christle}}, \bibinfo
  {author} {\bibfnamefont {G.}~\bibnamefont {Burkard}}, \bibinfo {author}
  {\bibfnamefont {F.~J.}\ \bibnamefont {Heremans}}, \bibinfo {author}
  {\bibfnamefont {L.~C.}\ \bibnamefont {Bassett}}, \ and\ \bibinfo {author}
  {\bibfnamefont {D.~D.}\ \bibnamefont {Awschalom}},\ }\href@noop {} {\bibfield
   {journal} {\bibinfo  {journal} {Proceedings of the National Academy of
  Sciences}\ }\textbf {\bibinfo {volume} {110}},\ \bibinfo {pages} {7595}
  (\bibinfo {year} {2013})}\BibitemShut {NoStop}%
\bibitem [{\citenamefont {Rogers}\ \emph {et~al.}(2014)\citenamefont {Rogers},
  \citenamefont {Jahnke}, \citenamefont {Metsch}, \citenamefont {Sipahigil},
  \citenamefont {Binder}, \citenamefont {Teraji}, \citenamefont {Sumiya},
  \citenamefont {Isoya}, \citenamefont {Lukin}, \citenamefont {Hemmer} \emph
  {et~al.}}]{rogers2014}%
  \BibitemOpen
  \bibfield  {author} {\bibinfo {author} {\bibfnamefont {L.~J.}\ \bibnamefont
  {Rogers}}, \bibinfo {author} {\bibfnamefont {K.~D.}\ \bibnamefont {Jahnke}},
  \bibinfo {author} {\bibfnamefont {M.~H.}\ \bibnamefont {Metsch}}, \bibinfo
  {author} {\bibfnamefont {A.}~\bibnamefont {Sipahigil}}, \bibinfo {author}
  {\bibfnamefont {J.~M.}\ \bibnamefont {Binder}}, \bibinfo {author}
  {\bibfnamefont {T.}~\bibnamefont {Teraji}}, \bibinfo {author} {\bibfnamefont
  {H.}~\bibnamefont {Sumiya}}, \bibinfo {author} {\bibfnamefont
  {J.}~\bibnamefont {Isoya}}, \bibinfo {author} {\bibfnamefont {M.~D.}\
  \bibnamefont {Lukin}}, \bibinfo {author} {\bibfnamefont {P.}~\bibnamefont
  {Hemmer}},  \emph {et~al.},\ }\href@noop {} {\bibfield  {journal} {\bibinfo
  {journal} {Physical review letters}\ }\textbf {\bibinfo {volume} {113}},\
  \bibinfo {pages} {263602} (\bibinfo {year} {2014})}\BibitemShut {NoStop}%
\bibitem [{\citenamefont {Pingault}\ \emph {et~al.}(2014)\citenamefont
  {Pingault}, \citenamefont {Becker}, \citenamefont {Schulte}, \citenamefont
  {Arend}, \citenamefont {Hepp}, \citenamefont {Godde}, \citenamefont
  {Tartakovskii}, \citenamefont {Markham}, \citenamefont {Becher},\ and\
  \citenamefont {Atat{\"u}re}}]{pingault2014}%
  \BibitemOpen
  \bibfield  {author} {\bibinfo {author} {\bibfnamefont {B.}~\bibnamefont
  {Pingault}}, \bibinfo {author} {\bibfnamefont {J.~N.}\ \bibnamefont
  {Becker}}, \bibinfo {author} {\bibfnamefont {C.~H.}\ \bibnamefont {Schulte}},
  \bibinfo {author} {\bibfnamefont {C.}~\bibnamefont {Arend}}, \bibinfo
  {author} {\bibfnamefont {C.}~\bibnamefont {Hepp}}, \bibinfo {author}
  {\bibfnamefont {T.}~\bibnamefont {Godde}}, \bibinfo {author} {\bibfnamefont
  {A.~I.}\ \bibnamefont {Tartakovskii}}, \bibinfo {author} {\bibfnamefont
  {M.}~\bibnamefont {Markham}}, \bibinfo {author} {\bibfnamefont
  {C.}~\bibnamefont {Becher}}, \ and\ \bibinfo {author} {\bibfnamefont
  {M.}~\bibnamefont {Atat{\"u}re}},\ }\href@noop {} {\bibfield  {journal}
  {\bibinfo  {journal} {Physical review letters}\ }\textbf {\bibinfo {volume}
  {113}},\ \bibinfo {pages} {263601} (\bibinfo {year} {2014})}\BibitemShut
  {NoStop}%
\bibitem [{\citenamefont {Jamonneau}\ \emph {et~al.}(2016)\citenamefont
  {Jamonneau}, \citenamefont {H{\'e}tet}, \citenamefont {Dr{\'e}au},
  \citenamefont {Roch},\ and\ \citenamefont {Jacques}}]{jamonneau2016}%
  \BibitemOpen
  \bibfield  {author} {\bibinfo {author} {\bibfnamefont {P.}~\bibnamefont
  {Jamonneau}}, \bibinfo {author} {\bibfnamefont {G.}~\bibnamefont
  {H{\'e}tet}}, \bibinfo {author} {\bibfnamefont {A.}~\bibnamefont
  {Dr{\'e}au}}, \bibinfo {author} {\bibfnamefont {J.-F.}\ \bibnamefont {Roch}},
  \ and\ \bibinfo {author} {\bibfnamefont {V.}~\bibnamefont {Jacques}},\
  }\href@noop {} {\bibfield  {journal} {\bibinfo  {journal} {Physical review
  letters}\ }\textbf {\bibinfo {volume} {116}},\ \bibinfo {pages} {043603}
  (\bibinfo {year} {2016})}\BibitemShut {NoStop}%
\bibitem [{\citenamefont {Smeltzer}\ \emph {et~al.}(2009)\citenamefont
  {Smeltzer}, \citenamefont {McIntyre},\ and\ \citenamefont
  {Childress}}]{smeltzer2009}%
  \BibitemOpen
  \bibfield  {author} {\bibinfo {author} {\bibfnamefont {B.}~\bibnamefont
  {Smeltzer}}, \bibinfo {author} {\bibfnamefont {J.}~\bibnamefont {McIntyre}},
  \ and\ \bibinfo {author} {\bibfnamefont {L.}~\bibnamefont {Childress}},\
  }\href@noop {} {\bibfield  {journal} {\bibinfo  {journal} {Physical Review
  A}\ }\textbf {\bibinfo {volume} {80}},\ \bibinfo {pages} {050302} (\bibinfo
  {year} {2009})}\BibitemShut {NoStop}%
\bibitem [{\citenamefont {Englund}\ \emph {et~al.}(2010)\citenamefont
  {Englund}, \citenamefont {Shields}, \citenamefont {Rivoire}, \citenamefont
  {Hatami}, \citenamefont {Vuckovic}, \citenamefont {Park},\ and\ \citenamefont
  {Lukin}}]{englund2010}%
  \BibitemOpen
  \bibfield  {author} {\bibinfo {author} {\bibfnamefont {D.}~\bibnamefont
  {Englund}}, \bibinfo {author} {\bibfnamefont {B.}~\bibnamefont {Shields}},
  \bibinfo {author} {\bibfnamefont {K.}~\bibnamefont {Rivoire}}, \bibinfo
  {author} {\bibfnamefont {F.}~\bibnamefont {Hatami}}, \bibinfo {author}
  {\bibfnamefont {J.}~\bibnamefont {Vuckovic}}, \bibinfo {author}
  {\bibfnamefont {H.}~\bibnamefont {Park}}, \ and\ \bibinfo {author}
  {\bibfnamefont {M.~D.}\ \bibnamefont {Lukin}},\ }\href@noop {} {\bibfield
  {journal} {\bibinfo  {journal} {Nano letters}\ }\textbf {\bibinfo {volume}
  {10}},\ \bibinfo {pages} {3922} (\bibinfo {year} {2010})}\BibitemShut
  {NoStop}%
\bibitem [{\citenamefont {De~Lange}\ \emph {et~al.}(2012)\citenamefont
  {De~Lange}, \citenamefont {Van Der~Sar}, \citenamefont {Blok}, \citenamefont
  {Wang}, \citenamefont {Dobrovitski},\ and\ \citenamefont {Hanson}}]{de2012}%
  \BibitemOpen
  \bibfield  {author} {\bibinfo {author} {\bibfnamefont {G.}~\bibnamefont
  {De~Lange}}, \bibinfo {author} {\bibfnamefont {T.}~\bibnamefont {Van
  Der~Sar}}, \bibinfo {author} {\bibfnamefont {M.}~\bibnamefont {Blok}},
  \bibinfo {author} {\bibfnamefont {Z.-H.}\ \bibnamefont {Wang}}, \bibinfo
  {author} {\bibfnamefont {V.}~\bibnamefont {Dobrovitski}}, \ and\ \bibinfo
  {author} {\bibfnamefont {R.}~\bibnamefont {Hanson}},\ }\href@noop {}
  {\bibfield  {journal} {\bibinfo  {journal} {Scientific reports}\ }\textbf
  {\bibinfo {volume} {2}},\ \bibinfo {pages} {1} (\bibinfo {year}
  {2012})}\BibitemShut {NoStop}%
\bibitem [{\citenamefont {Belthangady}\ \emph {et~al.}(2013)\citenamefont
  {Belthangady}, \citenamefont {Bar-Gill}, \citenamefont {Pham}, \citenamefont
  {Arai}, \citenamefont {Le~Sage}, \citenamefont {Cappellaro},\ and\
  \citenamefont {Walsworth}}]{belthangady2013}%
  \BibitemOpen
  \bibfield  {author} {\bibinfo {author} {\bibfnamefont {C.}~\bibnamefont
  {Belthangady}}, \bibinfo {author} {\bibfnamefont {N.}~\bibnamefont
  {Bar-Gill}}, \bibinfo {author} {\bibfnamefont {L.~M.}\ \bibnamefont {Pham}},
  \bibinfo {author} {\bibfnamefont {K.}~\bibnamefont {Arai}}, \bibinfo {author}
  {\bibfnamefont {D.}~\bibnamefont {Le~Sage}}, \bibinfo {author} {\bibfnamefont
  {P.}~\bibnamefont {Cappellaro}}, \ and\ \bibinfo {author} {\bibfnamefont
  {R.~L.}\ \bibnamefont {Walsworth}},\ }\href@noop {} {\bibfield  {journal}
  {\bibinfo  {journal} {Physical review letters}\ }\textbf {\bibinfo {volume}
  {110}},\ \bibinfo {pages} {157601} (\bibinfo {year} {2013})}\BibitemShut
  {NoStop}%
\bibitem [{\citenamefont {Childress}\ and\ \citenamefont
  {Hanson}(2013)}]{childress2013}%
  \BibitemOpen
  \bibfield  {author} {\bibinfo {author} {\bibfnamefont {L.}~\bibnamefont
  {Childress}}\ and\ \bibinfo {author} {\bibfnamefont {R.}~\bibnamefont
  {Hanson}},\ }\href@noop {} {\bibfield  {journal} {\bibinfo  {journal} {MRS
  bulletin}\ }\textbf {\bibinfo {volume} {38}},\ \bibinfo {pages} {134}
  (\bibinfo {year} {2013})}\BibitemShut {NoStop}%
\bibitem [{\citenamefont {Northup}\ and\ \citenamefont
  {Blatt}(2014)}]{northup2014}%
  \BibitemOpen
  \bibfield  {author} {\bibinfo {author} {\bibfnamefont {T.~E.}\ \bibnamefont
  {Northup}}\ and\ \bibinfo {author} {\bibfnamefont {R.}~\bibnamefont
  {Blatt}},\ }\href@noop {} {\bibfield  {journal} {\bibinfo  {journal} {Nature
  photonics}\ }\textbf {\bibinfo {volume} {8}},\ \bibinfo {pages} {356}
  (\bibinfo {year} {2014})}\BibitemShut {NoStop}%
\bibitem [{\citenamefont {Rohr}\ \emph {et~al.}(2014)\citenamefont {Rohr},
  \citenamefont {Dupont-Ferrier}, \citenamefont {Pigeau}, \citenamefont
  {Verlot}, \citenamefont {Jacques},\ and\ \citenamefont {Arcizet}}]{rohr2014}%
  \BibitemOpen
  \bibfield  {author} {\bibinfo {author} {\bibfnamefont {S.}~\bibnamefont
  {Rohr}}, \bibinfo {author} {\bibfnamefont {E.}~\bibnamefont
  {Dupont-Ferrier}}, \bibinfo {author} {\bibfnamefont {B.}~\bibnamefont
  {Pigeau}}, \bibinfo {author} {\bibfnamefont {P.}~\bibnamefont {Verlot}},
  \bibinfo {author} {\bibfnamefont {V.}~\bibnamefont {Jacques}}, \ and\
  \bibinfo {author} {\bibfnamefont {O.}~\bibnamefont {Arcizet}},\ }\href@noop
  {} {\bibfield  {journal} {\bibinfo  {journal} {Physical review letters}\
  }\textbf {\bibinfo {volume} {112}},\ \bibinfo {pages} {010502} (\bibinfo
  {year} {2014})}\BibitemShut {NoStop}%
\bibitem [{\citenamefont {Golter}\ \emph {et~al.}(2014)\citenamefont {Golter},
  \citenamefont {Baldwin},\ and\ \citenamefont {Wang}}]{golter2014}%
  \BibitemOpen
  \bibfield  {author} {\bibinfo {author} {\bibfnamefont {D.~A.}\ \bibnamefont
  {Golter}}, \bibinfo {author} {\bibfnamefont {T.~K.}\ \bibnamefont {Baldwin}},
  \ and\ \bibinfo {author} {\bibfnamefont {H.}~\bibnamefont {Wang}},\
  }\href@noop {} {\bibfield  {journal} {\bibinfo  {journal} {Physical review
  letters}\ }\textbf {\bibinfo {volume} {113}},\ \bibinfo {pages} {237601}
  (\bibinfo {year} {2014})}\BibitemShut {NoStop}%
\bibitem [{\citenamefont {Lee}\ \emph {et~al.}(2017)\citenamefont {Lee},
  \citenamefont {Lee}, \citenamefont {Cady}, \citenamefont {Ovartchaiyapong},\
  and\ \citenamefont {{Bleszynski Jayich}}}]{lee2017}%
  \BibitemOpen
  \bibfield  {author} {\bibinfo {author} {\bibfnamefont {D.}~\bibnamefont
  {Lee}}, \bibinfo {author} {\bibfnamefont {K.~W.}\ \bibnamefont {Lee}},
  \bibinfo {author} {\bibfnamefont {J.~V.}\ \bibnamefont {Cady}}, \bibinfo
  {author} {\bibfnamefont {P.}~\bibnamefont {Ovartchaiyapong}}, \ and\ \bibinfo
  {author} {\bibfnamefont {A.}~\bibnamefont {{Bleszynski Jayich}}},\
  }\href@noop {} {\bibfield  {journal} {\bibinfo  {journal} {Journal of
  Optics}\ }\textbf {\bibinfo {volume} {19}},\ \bibinfo {pages} {033001}
  (\bibinfo {year} {2017})}\BibitemShut {NoStop}%
\bibitem [{\citenamefont {Gu}\ \emph {et~al.}(2017)\citenamefont {Gu},
  \citenamefont {Kockum}, \citenamefont {Miranowicz}, \citenamefont {Liu},\
  and\ \citenamefont {Nori}}]{gu2017}%
  \BibitemOpen
  \bibfield  {author} {\bibinfo {author} {\bibfnamefont {X.}~\bibnamefont
  {Gu}}, \bibinfo {author} {\bibfnamefont {A.~F.}\ \bibnamefont {Kockum}},
  \bibinfo {author} {\bibfnamefont {A.}~\bibnamefont {Miranowicz}}, \bibinfo
  {author} {\bibfnamefont {Y.-x.}\ \bibnamefont {Liu}}, \ and\ \bibinfo
  {author} {\bibfnamefont {F.}~\bibnamefont {Nori}},\ }\href@noop {} {\bibfield
   {journal} {\bibinfo  {journal} {Physics Reports}\ }\textbf {\bibinfo
  {volume} {718}},\ \bibinfo {pages} {1} (\bibinfo {year} {2017})}\BibitemShut
  {NoStop}%
\bibitem [{\citenamefont {Choi}\ \emph {et~al.}(2017)\citenamefont {Choi},
  \citenamefont {Choi}, \citenamefont {Kucsko}, \citenamefont {Maurer},
  \citenamefont {Shields}, \citenamefont {Sumiya}, \citenamefont {Onoda},
  \citenamefont {Isoya}, \citenamefont {Demler}, \citenamefont {Jelezko} \emph
  {et~al.}}]{choi2017}%
  \BibitemOpen
  \bibfield  {author} {\bibinfo {author} {\bibfnamefont {J.}~\bibnamefont
  {Choi}}, \bibinfo {author} {\bibfnamefont {S.}~\bibnamefont {Choi}}, \bibinfo
  {author} {\bibfnamefont {G.}~\bibnamefont {Kucsko}}, \bibinfo {author}
  {\bibfnamefont {P.~C.}\ \bibnamefont {Maurer}}, \bibinfo {author}
  {\bibfnamefont {B.~J.}\ \bibnamefont {Shields}}, \bibinfo {author}
  {\bibfnamefont {H.}~\bibnamefont {Sumiya}}, \bibinfo {author} {\bibfnamefont
  {S.}~\bibnamefont {Onoda}}, \bibinfo {author} {\bibfnamefont
  {J.}~\bibnamefont {Isoya}}, \bibinfo {author} {\bibfnamefont
  {E.}~\bibnamefont {Demler}}, \bibinfo {author} {\bibfnamefont
  {F.}~\bibnamefont {Jelezko}},  \emph {et~al.},\ }\href@noop {} {\bibfield
  {journal} {\bibinfo  {journal} {Physical review letters}\ }\textbf {\bibinfo
  {volume} {118}},\ \bibinfo {pages} {093601} (\bibinfo {year}
  {2017})}\BibitemShut {NoStop}%
\bibitem [{\citenamefont {Kucsko}\ \emph {et~al.}(2018)\citenamefont {Kucsko},
  \citenamefont {Choi}, \citenamefont {Choi}, \citenamefont {Maurer},
  \citenamefont {Zhou}, \citenamefont {Landig}, \citenamefont {Sumiya},
  \citenamefont {Onoda}, \citenamefont {Isoya}, \citenamefont {Jelezko} \emph
  {et~al.}}]{kucsko2018}%
  \BibitemOpen
  \bibfield  {author} {\bibinfo {author} {\bibfnamefont {G.}~\bibnamefont
  {Kucsko}}, \bibinfo {author} {\bibfnamefont {S.}~\bibnamefont {Choi}},
  \bibinfo {author} {\bibfnamefont {J.}~\bibnamefont {Choi}}, \bibinfo {author}
  {\bibfnamefont {P.~C.}\ \bibnamefont {Maurer}}, \bibinfo {author}
  {\bibfnamefont {H.}~\bibnamefont {Zhou}}, \bibinfo {author} {\bibfnamefont
  {R.}~\bibnamefont {Landig}}, \bibinfo {author} {\bibfnamefont
  {H.}~\bibnamefont {Sumiya}}, \bibinfo {author} {\bibfnamefont
  {S.}~\bibnamefont {Onoda}}, \bibinfo {author} {\bibfnamefont
  {J.}~\bibnamefont {Isoya}}, \bibinfo {author} {\bibfnamefont
  {F.}~\bibnamefont {Jelezko}},  \emph {et~al.},\ }\href@noop {} {\bibfield
  {journal} {\bibinfo  {journal} {Physical Review Letters}\ }\textbf {\bibinfo
  {volume} {121}},\ \bibinfo {pages} {023601} (\bibinfo {year}
  {2018})}\BibitemShut {NoStop}%
\bibitem [{\citenamefont {Astner}\ \emph {et~al.}(2018)\citenamefont {Astner},
  \citenamefont {Gugler}, \citenamefont {Angerer}, \citenamefont {Wald},
  \citenamefont {Putz}, \citenamefont {Mauser}, \citenamefont {Trupke},
  \citenamefont {Sumiya}, \citenamefont {Onoda}, \citenamefont {Isoya} \emph
  {et~al.}}]{astner2018}%
  \BibitemOpen
  \bibfield  {author} {\bibinfo {author} {\bibfnamefont {T.}~\bibnamefont
  {Astner}}, \bibinfo {author} {\bibfnamefont {J.}~\bibnamefont {Gugler}},
  \bibinfo {author} {\bibfnamefont {A.}~\bibnamefont {Angerer}}, \bibinfo
  {author} {\bibfnamefont {S.}~\bibnamefont {Wald}}, \bibinfo {author}
  {\bibfnamefont {S.}~\bibnamefont {Putz}}, \bibinfo {author} {\bibfnamefont
  {N.~J.}\ \bibnamefont {Mauser}}, \bibinfo {author} {\bibfnamefont
  {M.}~\bibnamefont {Trupke}}, \bibinfo {author} {\bibfnamefont
  {H.}~\bibnamefont {Sumiya}}, \bibinfo {author} {\bibfnamefont
  {S.}~\bibnamefont {Onoda}}, \bibinfo {author} {\bibfnamefont
  {J.}~\bibnamefont {Isoya}},  \emph {et~al.},\ }\href@noop {} {\bibfield
  {journal} {\bibinfo  {journal} {Nature materials}\ }\textbf {\bibinfo
  {volume} {17}},\ \bibinfo {pages} {313} (\bibinfo {year} {2018})}\BibitemShut
  {NoStop}%
\bibitem [{\citenamefont {Bauch}\ \emph {et~al.}(2018)\citenamefont {Bauch},
  \citenamefont {Hart}, \citenamefont {Schloss}, \citenamefont {Turner},
  \citenamefont {Barry}, \citenamefont {Kehayias}, \citenamefont {Singh},\ and\
  \citenamefont {Walsworth}}]{bauch2018}%
  \BibitemOpen
  \bibfield  {author} {\bibinfo {author} {\bibfnamefont {E.}~\bibnamefont
  {Bauch}}, \bibinfo {author} {\bibfnamefont {C.~A.}\ \bibnamefont {Hart}},
  \bibinfo {author} {\bibfnamefont {J.~M.}\ \bibnamefont {Schloss}}, \bibinfo
  {author} {\bibfnamefont {M.~J.}\ \bibnamefont {Turner}}, \bibinfo {author}
  {\bibfnamefont {J.~F.}\ \bibnamefont {Barry}}, \bibinfo {author}
  {\bibfnamefont {P.}~\bibnamefont {Kehayias}}, \bibinfo {author}
  {\bibfnamefont {S.}~\bibnamefont {Singh}}, \ and\ \bibinfo {author}
  {\bibfnamefont {R.~L.}\ \bibnamefont {Walsworth}},\ }\href@noop {} {\bibfield
   {journal} {\bibinfo  {journal} {Physical Review X}\ }\textbf {\bibinfo
  {volume} {8}},\ \bibinfo {pages} {031025} (\bibinfo {year}
  {2018})}\BibitemShut {NoStop}%
\bibitem [{\citenamefont {Barry}\ \emph {et~al.}(2020)\citenamefont {Barry},
  \citenamefont {Schloss}, \citenamefont {Bauch}, \citenamefont {Turner},
  \citenamefont {Hart}, \citenamefont {Pham},\ and\ \citenamefont
  {Walsworth}}]{barry2020}%
  \BibitemOpen
  \bibfield  {author} {\bibinfo {author} {\bibfnamefont {J.~F.}\ \bibnamefont
  {Barry}}, \bibinfo {author} {\bibfnamefont {J.~M.}\ \bibnamefont {Schloss}},
  \bibinfo {author} {\bibfnamefont {E.}~\bibnamefont {Bauch}}, \bibinfo
  {author} {\bibfnamefont {M.~J.}\ \bibnamefont {Turner}}, \bibinfo {author}
  {\bibfnamefont {C.~A.}\ \bibnamefont {Hart}}, \bibinfo {author}
  {\bibfnamefont {L.~M.}\ \bibnamefont {Pham}}, \ and\ \bibinfo {author}
  {\bibfnamefont {R.~L.}\ \bibnamefont {Walsworth}},\ }\href@noop {} {\bibfield
   {journal} {\bibinfo  {journal} {Reviews of Modern Physics}\ }\textbf
  {\bibinfo {volume} {92}},\ \bibinfo {pages} {015004} (\bibinfo {year}
  {2020})}\BibitemShut {NoStop}%
\bibitem [{\citenamefont {Kehayias}\ \emph {et~al.}(2014)\citenamefont
  {Kehayias}, \citenamefont {Mr{\'o}zek}, \citenamefont {Acosta}, \citenamefont
  {Jarmola}, \citenamefont {Rudnicki}, \citenamefont {Folman}, \citenamefont
  {Gawlik},\ and\ \citenamefont {Budker}}]{kehayias2014}%
  \BibitemOpen
  \bibfield  {author} {\bibinfo {author} {\bibfnamefont {P.}~\bibnamefont
  {Kehayias}}, \bibinfo {author} {\bibfnamefont {M.}~\bibnamefont
  {Mr{\'o}zek}}, \bibinfo {author} {\bibfnamefont {V.~M.}\ \bibnamefont
  {Acosta}}, \bibinfo {author} {\bibfnamefont {A.}~\bibnamefont {Jarmola}},
  \bibinfo {author} {\bibfnamefont {D.}~\bibnamefont {Rudnicki}}, \bibinfo
  {author} {\bibfnamefont {R.}~\bibnamefont {Folman}}, \bibinfo {author}
  {\bibfnamefont {W.}~\bibnamefont {Gawlik}}, \ and\ \bibinfo {author}
  {\bibfnamefont {D.}~\bibnamefont {Budker}},\ }\href@noop {} {\bibfield
  {journal} {\bibinfo  {journal} {Physical Review B}\ }\textbf {\bibinfo
  {volume} {89}},\ \bibinfo {pages} {245202} (\bibinfo {year}
  {2014})}\BibitemShut {NoStop}%
\bibitem [{\citenamefont {Putz}\ \emph {et~al.}(2017)\citenamefont {Putz},
  \citenamefont {Angerer}, \citenamefont {Krimer}, \citenamefont {Glattauer},
  \citenamefont {Munro}, \citenamefont {Rotter}, \citenamefont {Schmiedmayer},\
  and\ \citenamefont {Majer}}]{putz2017}%
  \BibitemOpen
  \bibfield  {author} {\bibinfo {author} {\bibfnamefont {S.}~\bibnamefont
  {Putz}}, \bibinfo {author} {\bibfnamefont {A.}~\bibnamefont {Angerer}},
  \bibinfo {author} {\bibfnamefont {D.~O.}\ \bibnamefont {Krimer}}, \bibinfo
  {author} {\bibfnamefont {R.}~\bibnamefont {Glattauer}}, \bibinfo {author}
  {\bibfnamefont {W.~J.}\ \bibnamefont {Munro}}, \bibinfo {author}
  {\bibfnamefont {S.}~\bibnamefont {Rotter}}, \bibinfo {author} {\bibfnamefont
  {J.}~\bibnamefont {Schmiedmayer}}, \ and\ \bibinfo {author} {\bibfnamefont
  {J.}~\bibnamefont {Majer}},\ }\href@noop {} {\bibfield  {journal} {\bibinfo
  {journal} {Nature Photonics}\ }\textbf {\bibinfo {volume} {11}},\ \bibinfo
  {pages} {36} (\bibinfo {year} {2017})}\BibitemShut {NoStop}%
\bibitem [{\citenamefont {Soltamov}\ \emph {et~al.}(2019)\citenamefont
  {Soltamov}, \citenamefont {Kasper}, \citenamefont {Poshakinskiy},
  \citenamefont {Anisimov}, \citenamefont {Mokhov}, \citenamefont {Sperlich},
  \citenamefont {Tarasenko}, \citenamefont {Baranov}, \citenamefont
  {Astakhov},\ and\ \citenamefont {Dyakonov}}]{soltamov2019}%
  \BibitemOpen
  \bibfield  {author} {\bibinfo {author} {\bibfnamefont {V.~A.}\ \bibnamefont
  {Soltamov}}, \bibinfo {author} {\bibfnamefont {C.}~\bibnamefont {Kasper}},
  \bibinfo {author} {\bibfnamefont {A.~V.}\ \bibnamefont {Poshakinskiy}},
  \bibinfo {author} {\bibfnamefont {A.~N.}\ \bibnamefont {Anisimov}}, \bibinfo
  {author} {\bibfnamefont {E.~N.}\ \bibnamefont {Mokhov}}, \bibinfo {author}
  {\bibfnamefont {A.}~\bibnamefont {Sperlich}}, \bibinfo {author}
  {\bibfnamefont {S.~A.}\ \bibnamefont {Tarasenko}}, \bibinfo {author}
  {\bibfnamefont {P.~G.}\ \bibnamefont {Baranov}}, \bibinfo {author}
  {\bibfnamefont {G.~V.}\ \bibnamefont {Astakhov}}, \ and\ \bibinfo {author}
  {\bibfnamefont {V.}~\bibnamefont {Dyakonov}},\ }\href@noop {} {\bibfield
  {journal} {\bibinfo  {journal} {Nature communications}\ }\textbf {\bibinfo
  {volume} {10}},\ \bibinfo {pages} {1} (\bibinfo {year} {2019})}\BibitemShut
  {NoStop}%
\bibitem [{\citenamefont {Mr{\'o}zek}\ \emph {et~al.}(2016)\citenamefont
  {Mr{\'o}zek}, \citenamefont {Wojciechowski}, \citenamefont {Rudnicki},
  \citenamefont {Zachorowski}, \citenamefont {Kehayias}, \citenamefont
  {Budker},\ and\ \citenamefont {Gawlik}}]{mrozek2016}%
  \BibitemOpen
  \bibfield  {author} {\bibinfo {author} {\bibfnamefont {M.}~\bibnamefont
  {Mr{\'o}zek}}, \bibinfo {author} {\bibfnamefont {A.~M.}\ \bibnamefont
  {Wojciechowski}}, \bibinfo {author} {\bibfnamefont {D.}~\bibnamefont
  {Rudnicki}}, \bibinfo {author} {\bibfnamefont {J.}~\bibnamefont
  {Zachorowski}}, \bibinfo {author} {\bibfnamefont {P.}~\bibnamefont
  {Kehayias}}, \bibinfo {author} {\bibfnamefont {D.}~\bibnamefont {Budker}}, \
  and\ \bibinfo {author} {\bibfnamefont {W.}~\bibnamefont {Gawlik}},\
  }\href@noop {} {\bibfield  {journal} {\bibinfo  {journal} {Physical Review
  B}\ }\textbf {\bibinfo {volume} {94}},\ \bibinfo {pages} {035204} (\bibinfo
  {year} {2016})}\BibitemShut {NoStop}%
\bibitem [{\citenamefont {Baklanov}\ and\ \citenamefont
  {Chebotaev}(1972)}]{baklanov1972}%
  \BibitemOpen
  \bibfield  {author} {\bibinfo {author} {\bibfnamefont {E.~V.}\ \bibnamefont
  {Baklanov}}\ and\ \bibinfo {author} {\bibfnamefont {V.~P.}\ \bibnamefont
  {Chebotaev}},\ }\href@noop {} {\bibfield  {journal} {\bibinfo  {journal}
  {Sov. Phys. JETP}\ }\textbf {\bibinfo {volume} {34}},\ \bibinfo {pages} {287}
  (\bibinfo {year} {1972})}\BibitemShut {NoStop}%
\bibitem [{\citenamefont {Sargent~III}(1978)}]{sargent1978}%
  \BibitemOpen
  \bibfield  {author} {\bibinfo {author} {\bibfnamefont {M.}~\bibnamefont
  {Sargent~III}},\ }\href@noop {} {\bibfield  {journal} {\bibinfo  {journal}
  {Physics Reports}\ }\textbf {\bibinfo {volume} {43}},\ \bibinfo {pages} {223}
  (\bibinfo {year} {1978})}\BibitemShut {NoStop}%
\bibitem [{\citenamefont {Laupr{\^e}tre}\ \emph {et~al.}(2012)\citenamefont
  {Laupr{\^e}tre}, \citenamefont {Kumar}, \citenamefont {Berger}, \citenamefont
  {Faoro}, \citenamefont {Ghosh}, \citenamefont {Bretenaker},\ and\
  \citenamefont {Goldfarb}}]{laupretre2012}%
  \BibitemOpen
  \bibfield  {author} {\bibinfo {author} {\bibfnamefont {T.}~\bibnamefont
  {Laupr{\^e}tre}}, \bibinfo {author} {\bibfnamefont {S.}~\bibnamefont
  {Kumar}}, \bibinfo {author} {\bibfnamefont {P.}~\bibnamefont {Berger}},
  \bibinfo {author} {\bibfnamefont {R.}~\bibnamefont {Faoro}}, \bibinfo
  {author} {\bibfnamefont {R.}~\bibnamefont {Ghosh}}, \bibinfo {author}
  {\bibfnamefont {F.}~\bibnamefont {Bretenaker}}, \ and\ \bibinfo {author}
  {\bibfnamefont {F.}~\bibnamefont {Goldfarb}},\ }\href@noop {} {\bibfield
  {journal} {\bibinfo  {journal} {Physical Review A}\ }\textbf {\bibinfo
  {volume} {85}},\ \bibinfo {pages} {051805} (\bibinfo {year}
  {2012})}\BibitemShut {NoStop}%
\bibitem [{\citenamefont {El-Ella}\ \emph {et~al.}(2019)\citenamefont
  {El-Ella}, \citenamefont {Huck},\ and\ \citenamefont {Andersen}}]{ella2019}%
  \BibitemOpen
  \bibfield  {author} {\bibinfo {author} {\bibfnamefont {H.~A.~R.}\
  \bibnamefont {El-Ella}}, \bibinfo {author} {\bibfnamefont {A.}~\bibnamefont
  {Huck}}, \ and\ \bibinfo {author} {\bibfnamefont {U.~L.}\ \bibnamefont
  {Andersen}},\ }\href@noop {} {\bibfield  {journal} {\bibinfo  {journal}
  {Physical Review B}\ }\textbf {\bibinfo {volume} {100}},\ \bibinfo {pages}
  {214407} (\bibinfo {year} {2019})}\BibitemShut {NoStop}%
\bibitem [{sup()}]{suplement}%
  \BibitemOpen
  \href@noop {} {}\bibinfo {note} {{See Supplemental Information to this
  Letter}}\BibitemShut {NoStop}%
\bibitem [{\citenamefont {Mr{\'o}zek}\ \emph {et~al.}(2015)\citenamefont
  {Mr{\'o}zek}, \citenamefont {Rudnicki}, \citenamefont {Kehayias},
  \citenamefont {Jarmola}, \citenamefont {Budker},\ and\ \citenamefont
  {Gawlik}}]{mrozek2015}%
  \BibitemOpen
  \bibfield  {author} {\bibinfo {author} {\bibfnamefont {M.}~\bibnamefont
  {Mr{\'o}zek}}, \bibinfo {author} {\bibfnamefont {D.}~\bibnamefont
  {Rudnicki}}, \bibinfo {author} {\bibfnamefont {P.}~\bibnamefont {Kehayias}},
  \bibinfo {author} {\bibfnamefont {A.}~\bibnamefont {Jarmola}}, \bibinfo
  {author} {\bibfnamefont {D.}~\bibnamefont {Budker}}, \ and\ \bibinfo {author}
  {\bibfnamefont {W.}~\bibnamefont {Gawlik}},\ }\href@noop {} {\bibfield
  {journal} {\bibinfo  {journal} {EPJ Quantum Technology}\ }\textbf {\bibinfo
  {volume} {2}},\ \bibinfo {pages} {1} (\bibinfo {year} {2015})}\BibitemShut
  {NoStop}%
\bibitem [{\citenamefont {Dr{\'e}au}\ \emph {et~al.}(2011)\citenamefont
  {Dr{\'e}au}, \citenamefont {Lesik}, \citenamefont {Rondin}, \citenamefont
  {Spinicelli}, \citenamefont {Arcizet}, \citenamefont {Roch},\ and\
  \citenamefont {Jacques}}]{dreau2011}%
  \BibitemOpen
  \bibfield  {author} {\bibinfo {author} {\bibfnamefont {A.}~\bibnamefont
  {Dr{\'e}au}}, \bibinfo {author} {\bibfnamefont {M.}~\bibnamefont {Lesik}},
  \bibinfo {author} {\bibfnamefont {L.}~\bibnamefont {Rondin}}, \bibinfo
  {author} {\bibfnamefont {P.}~\bibnamefont {Spinicelli}}, \bibinfo {author}
  {\bibfnamefont {O.}~\bibnamefont {Arcizet}}, \bibinfo {author} {\bibfnamefont
  {J.-F.}\ \bibnamefont {Roch}}, \ and\ \bibinfo {author} {\bibfnamefont
  {V.}~\bibnamefont {Jacques}},\ }\href@noop {} {\bibfield  {journal} {\bibinfo
   {journal} {Physical Review B}\ }\textbf {\bibinfo {volume} {84}},\ \bibinfo
  {pages} {195204} (\bibinfo {year} {2011})},\ \bibinfo {note} {the observed
  independence of w0 on light power should not be confused with the effect
  reported in: A. Dr{\'e}au, M. Lesik, L. Rondin, P. Spinicelli, O. Arcizet,
  J-F. Roch, and V. Jacques, Physical Review B \textbf{84}, 195204
  (2011).}\BibitemShut {Stop}%
\bibitem [{\citenamefont {Jensen}\ \emph {et~al.}(2013)\citenamefont {Jensen},
  \citenamefont {Acosta}, \citenamefont {Jarmola},\ and\ \citenamefont
  {Budker}}]{jensen2013}%
  \BibitemOpen
  \bibfield  {author} {\bibinfo {author} {\bibfnamefont {K.}~\bibnamefont
  {Jensen}}, \bibinfo {author} {\bibfnamefont {V.~M.}\ \bibnamefont {Acosta}},
  \bibinfo {author} {\bibfnamefont {A.}~\bibnamefont {Jarmola}}, \ and\
  \bibinfo {author} {\bibfnamefont {D.}~\bibnamefont {Budker}},\ }\href@noop {}
  {\bibfield  {journal} {\bibinfo  {journal} {Physical Review B}\ }\textbf
  {\bibinfo {volume} {87}},\ \bibinfo {pages} {014115} (\bibinfo {year}
  {2013})}\BibitemShut {NoStop}%
\end{thebibliography}
%

\end{document}


\title{Spin State Dynamics in a Bichromatic Microwave Field: Role of Bright and Dark States in coupling with Reservoir {-- Supplemental Materials}}
\author{Wojciech Gawlik}
\author{Piotr Olczykowski}
\author{Mariusz Mr\'ozek}
\author{Adam M. Wojciechowski}
\affiliation{Institute  of  Physics,  Jagiellonian  University, {\L}ojasiewicza  11,  30-348  Krak\'ow,  Poland}


\let\stdsettings\newpage
\let\newpage\relax
\maketitle
\let\newpage\stdsettings

\section{Introduction}
The analysis is based on a simplified two-level system and is performed with the density matrix formalism. The quantum master equation yields:
\begin{equation} \label{eq:alephmatrix}
\dfrac{d}{dt}\rho= -\frac{i}{\hbar}[H,\rho] - (\Gamma \rho),
\end{equation}
with the density matrix 
\[\rho = \begin{pmatrix} \pii & \pio \\ \poi &\poo\\ \end{pmatrix}\]
and the Hamiltonian defined in the frame rotating with frequency \(\omega_1\):
\[ H = \frac \hbar 2 \begin{pmatrix} -\omega & \Os+\Op e^{i\delta t} \\
\Os+ \Op e^{-i\delta t} & \omega \end{pmatrix}, \]
where \(\Os,\, \Op\) denote Rabi frequencies of two \(MW\)s which have their frequencies differing by \(\delta = \os-\op\) and \(\omega = \os-\omega_0\) represents detuning of \(\omega_1\) from the ODMR central frequency \(\omega_0\).

The part of the evolution generator responsible for relaxation and decoherence is independent of the choice of the rotating frame and defined as:
\[(\Gamma \rho) = \begin{pmatrix}\gamma_1(\pii-\pii^0)& \Gamma \pio \\
\Gamma \poi & \gamma_0(\poo-\poo^0)\\ \end{pmatrix},\]
where \(\rho_{kk}^0\) are equilibrium populations, \(\gamma_k=\gamma(1-(-1)^k \epsilon)\) denote the rates with which populations \(\rho_{kk}\) relax to equilibrium, and \(\Gamma\) is the relaxation rate of coherences \(\rho_{jk}\) (here and throughout the paper indices \(j,\,k = 0,\,1\)). Moreover, we assume that both \(\Gamma\) and \(\gamma\) are nonzero. Master equation (Eq.~\ref{eq:alephmatrix}) results then in the following set of equations: 
\begin{equation}\label{eq:aleph}
\begin{split}
\dpii &= - \gamma(1+\epsilon) (\pii-\pii^0)+ \frac{i}{2} (\Os+\Op e^{-i\delta t})\pio - \frac i2 (\Os+\Op e^{i\delta t})\poi,\\
\dpoo &= - \gamma(1-\epsilon) (\poo-\poo^0)-\frac{i}{2} (\Os+\Op e^{-i\delta t})\pio + \frac i2  (\Os+\Op e^{i\delta t})\poi ,\\
\dpio &= -(\Gamma - i \omega)\pio + \frac i2  (\Os+\Op e^{i\delta t}) (\pii-\poo),\\
\dpoi &= -(\Gamma + i \omega)\poi - \frac i2  (\Os+\Op e^{- i\delta t})(\pii-\poo).
\end{split}
\end{equation}

\section{Closed vs. open systems}
The trace of the density matrix represents the sum of individual populations. Since the master equation describes the evolution of the density matrix, its trace represents the time derivative of total probability stored in the system. For a closed system total probability is conserved, so \(\mathrm{Tr}\dot \rho=0\). On the contrary, the systems with \(\mathrm{Tr}\dot \rho\neq0\) are referred to as open ones since the probability flows in and out of the system from a reservoir of probability.
To examine the conditions of openness of the system we compute the trace of Eq.~\ref{eq:alephmatrix}: 
\[\dpii + \dpoo = -\gamma\Big((1+\epsilon)(\pii-\pii^0)+(1-\epsilon)(\poo-\poo^0)\Big).\]
Conservation of the total probability requires that the left-hand side of the above equation nullifies, i.e. \(\dpoo +\dpii =0\), which results in:
\[ \pii+\poo +\epsilon(\pii-\poo) = \pii^0+\poo^0 +\epsilon(\pii^0-\poo^0).\]
Without loss of generality, we can define a normalization condition: \(\pii^0+\poo^0=1\). For \(\epsilon=0\) there exists non-stationary solution for which the total probability is conserved with the normalization \(\pii+\poo=1\). On the contrary, for \(\epsilon\neq 0\) the set of equations:
\[ \left\{ \begin{array}{rcl}
\pii+\poo & = &1  \\
\epsilon(\pii-\poo) &= &\epsilon(\pii^0-\poo^0)
\end{array} \right. \]
has only stationary solution with \(\rho_{kk}=\rho_{kk}^0\). This solution, in turn, does not fulfill Equation \ref{eq:aleph} for \(\Os\neq 0\) or \(\Op\neq 0\) if only \(\delta\neq 0\) and \(\pii^0-\poo^0 \neq 0\).

To summarize, equality of relaxation rates \(\gamma_0=\gamma_1\) signifies a closed system with the normalization condition \(\pii+\poo=\pii^0+\poo^0\), whereas inequality \(\gamma_0\neq\gamma_1\) represented by the existence of asymmetry parameter \(\epsilon\neq0,\) requires opening of the system. In the open system probability is not conserved, flows in and out  of the system from a reservoir of probability. Further, our analysis shows that the openness of the system under consideration can be detected only for two \(MW\)s oscillating with different frequencies \(\omega_1\neq\omega_0\) and the system is prepared in such a manner that its equilibrium populations are not equal. i.e. \(\pii^0\neq\poo^0\).

This can be explained in the following way, for a closed system (\(\epsilon=0\)) the solution is restricted to a one-dimensional space (a line) with \(\pii+\poo=1\), on the contrary, for \(\epsilon\neq0\) the solution is not constrained by total probability, thus it penetrates the subset of a two-dimensional manifold parameterized by individual populations. Moreover, it is driven by \(MW\) field in the environment of the stationary solution, hence it can be decomposed into two evolving states identified by different relaxation rates and coupled by \(MW\) field. This very observation is sufficient to explain qualitatively the existence of two components of the resonance which reflects the dimensionality of the solution. As \(\mathrm{Tr}([H,\rho])=0\), this effect does not depend on the specific form of interaction and is general. 

\section{Adiabatic approximation}

By application of the Floquet theory we obtain that the solution exponentially approaches periodic function with fundamental period \(\frac{2 \pi}{\delta}\). Assuming that the time of measurement is much longer that the inverse of the Floquet exponentials, we expand the solution into the Fourier series, i.e.:
\[ \rho_{\alpha\beta}(t)=\sum_{k\in\ze} \tilde{\rho}_{\alpha\beta}(k)e^{ik\delta t} \quad \forall\, \alpha, \, \beta = 0,\,1.\]
The last two equations of Eq. \ref{eq:aleph} in the Fourier basis yield:
\begin{equation}\label{eq:gimel}\begin{split}
\tpio(k)&=\frac i2 \cdot \frac{\Os\Big(\tpii(k)-\tpoo(k)\Big)+\Op\Big(\tpii(k-1)-\tpoo(k-1)\Big)}{\Gamma- i \omega + ik \delta },\\
\tpoi(k)&=-\frac i2 \cdot \frac{\Os\Big(\tpii(k)-\tpoo(k)\Big)+\Op\Big(\tpii(k+1)-\tpoo(k+1)\Big)}{\Gamma+ i \omega + ik\delta }.
\end{split}\end{equation}
We use Eq. \ref{eq:gimel} to eliminate coherences from the first two equations \ref{eq:aleph} (for brevity we change the notation \(n_i:=\rho_{ii}\)):
\begin{equation}\begin{split} \label{eq:bet}
 \frac{i k \delta}{\gamma} \tni(k)= - (1+\epsilon)(\tni(k)-n_1^0 \delta_{k0}) - \frac 12 \mathcal{F}(k),\\
 \frac{ik \delta}{\gamma} \tno(k)= - (1-\epsilon)(\tno(k)-n_0^0 \delta_{k0})+ \frac 12 \mathcal{F}(k),
\end{split}\end{equation}
where:
 \[ \begin{split}
\mathcal{F}(k)=&\kh(k)\Big(\tni(k)-\tno(k)\Big) \\
&+\frac{\lh(k)}2 \Big(\tni(k+1)+\tni(k-1)-\tno(k+1)-\tno(k-1)\Big)\\
&+\frac{\mh(k)}2 \Big(\tni(k+1)-\tni(k-1)-\tno(k+1)+\tno(k-1)\Big),
 \end{split}\]
 \[\begin{split}
 \kh(k) & = \frac{\Os^2}{\gamma}\cdot  \frac{\Gamma+ik\delta }{(\Gamma+ik\delta)^2+(\os-\omega_0)^2}+\frac{\Op^2}{\gamma}\cdot \frac{\Gamma+ik\delta }{(\Gamma+ik\delta)^2+(\op-\omega_0)^2},\\
 \lh(k) &= \frac{\Os \Op}{\gamma} \cdot  \left( \frac{\Gamma+ik\delta }{(\Gamma+ik\delta)^2+(\os-\omega_0)^2}+\frac{\Gamma+ik\delta }{(\Gamma+ik\delta)^2+(\op-\omega_0)^2} \right),\\
 \mh(k) &= \frac{\Os \Op}{\gamma}  \cdot  \left( \frac{-(\os-\omega_0)}{(\Gamma+ik\delta)^2+(\os-\omega_0)^2}+\frac{\op-\omega_0}{(\Gamma+ik\delta)^2+(\op-\omega_0)^2} \right).
 \end{split}
 \]
 
Up to now, the calculations were exact, we will now make some simplifying assumptions:
\begin{itemize}
	\item we consider effects which occur for small detunings only:
	\[\frac{\delta}{\Gamma}\simeq 0,\]
	\item we assume that for very high harmonic numbers (\(n\simeq \frac{\Gamma}{\delta}\)) the amplitudes of the elements of the Fourier series are negligible:
	\[\tilde{n}_i(n)\simeq 0,\]
	\item we consider effects caused by burning a hole close to the center of the ODMR resonance:
	\[\frac{\omega_i-\omega_0}\Gamma\simeq 0,\]
	\item both MWs are taken as equally strong:
	\[\Os=\Op=: \Omega.\]
\end{itemize}
With these assumptions, the population dynamics  can be cast in a simple form of first-order inhomogeneous equations (presented in the main paper as Eqs. 1):
\begin{equation}\begin{split} \label{eq:dalet}
\frac{d}{d(\gamma t)} n_1= - (1+\epsilon)(n_1-n_1^0) - S\big(1 + \cos \delta t \big) (n_1-n_0),\\
\frac{d}{d(\gamma t)}n_0= - (1-\epsilon)(n_0-n_0^0) +  S\big(1 + \cos \delta t \big)  (n_1-n_0),
\end{split}\end{equation}
where:
 \[S = \frac{\Omega^2}{\gamma\Gamma}\Big(L _\Gamma (\omega_1-\omega_0)+L _{\Gamma} (\omega_2-\omega_0)\Big)\] 
and we use a normalized Lorentz function \(L_a(x)=\frac{a^2}{x^2+a^2}\).

\section{Liouville equation, eigenvalues, dark and bright states}
\subsection{Liouvillian expansion }
We decompose the Liouvillian into a sum of a time independent \(\mathcal{L}_0\) and time dependent \(\mathcal{L}_1(t)\), i.e.:
\[ \dot n   = - (\mathcal{L}_0+\mathcal{L}_1(t))\cdot n + n^0,\]
where:

\begin{equation}\label{eq:he}
\begin{split}
n & = \begin{pmatrix}n_1 \\n_0\end{pmatrix}, \quad
\mathcal{L}_0=\begin{pmatrix}1+S+\epsilon & -S \\-S & 1+S-\epsilon\end{pmatrix},\\
n^0&= \begin{pmatrix}(1+\epsilon)n_1^0 \\(1-\epsilon)n_0^0\end{pmatrix},
\quad \mathcal{L}_1(t)=S \cos \delta t \cdot \begin{pmatrix}1 & -1 \\-1 & 1\end{pmatrix},
\end{split} 
\end{equation}
which become Eqs. 2 in the main paper. Operators \( \mathcal{L}_0 \) and \( \mathcal{L}_1\) play significantly different roles in the dynamics of the system. The analysis of Eq. \ref{eq:he} is performed in three steps: first, to identify the states spanning the space of the  solutions interpreted as dressed populations and causing the emergence of two components of the composite resonance; second, to designate their time-dependent coupling; and last, to obtain a quantitative prediction on the widths of the resonance. To this aim we interpret  \(\mathcal{L}_0\) as dressing initial states resulting in creation of two superpositions which we denote \(\eta_0\) and \(\eta_1\) and \(\mathcal{L}_1\) as a time-dependent coupling between them. 	

\subsection{Time independent part}
By zeroing \(\mathcal{L}_1(t)\) in Eq. \ref{eq:he} we focus only on its time independent part. This part enables the definition of the reference system for analyzing the evolution of the full, time-dependent equation. A formal solution of Eq. \ref{eq:he}  can be written as

\begin{equation}\label{eq:he1}
n(t)=e^{-\mathcal{L}_0\cdot\gamma t}(n(0)-\bar n)+ \bar n,
\end{equation}
where point \(\bar n \colon= \mathcal{L}_0^{-1}n^0\) denotes the stationary solution:

\[\bar n  =\begin{pmatrix}n_1^0-\frac{S(1-\epsilon)}{1+2S-\epsilon^2}(n_1^0-n_0^0) \\ n_0^0 +\frac{S(1+\epsilon)}{1+2S-\epsilon^2}(n_1^0-n_0^0)\end{pmatrix}. \]
For the sake of identification of independently evolving states and their lifetimes, we diagonalize \(\mathcal{L}_0\)  with the help of the \(U_R\) matrix: \(U_R^{-1}\mathcal{L}_0 U_R=\mathrm{diag}(\lambda_1,\lambda_0)\),  with \(\lambda_1,\,\lambda_0\) being the eigenvalues of \(\mathcal{L}_0\). Straightforward computation yields:
\[U_R=\begin{pmatrix}\cos \theta & \sin \theta \\-\sin \theta & \cos \theta\end{pmatrix},\]
where we interpret \(\theta\) as the mixing angle between initial populations \(n_1\) i \(n_0\) and \(\theta\) satisfies the implicit equation:
\[\tan \theta = \sqrt{1+\left(\frac{\epsilon}{S}\right)^2}-\frac{\epsilon}{S}.\]

Eigenvalues of \(\mathcal{L}_0\) are:

\[\lambda_k=1+S-(-1)^k\sqrt{S^2+\epsilon^2} \quad \forall \, k=0,1.\]

The differences between the momentary populations and stationary ones, \(n(t)-\bar n^0\), span the vector space. The eigenvectors of \(\mathcal{L}_0\) resulting from the diagonalization can be interpreted as two independently evolving states \(\eta_1\) i \(\eta_0\) with coordinates \( \begin{pmatrix}\eta_1 \\ \eta_0 \end{pmatrix} = U_R^{-1} (n-\bar n)\), i.e.:

\begin{equation}\label{eq:chet}
\begin{split}
\eta_1 &=\cos \theta (n_1-\bar n_1)-\sin \theta (n_0-\bar n_0)\\
\eta_0 &=\sin \theta (n_1-\bar n_1)+\cos \theta (n_0-\bar n_0),
\end{split}
\end{equation}
which leads to the solution of the time-independent part of Eq. \ref{eq:he}:

\[ \eta(t) \colon = U_R^{-1}( n(t) - \bar n )= \begin{pmatrix}
e^{-\lambda_1 \cdot\gamma t}\eta_1(0) \\e^{-\lambda_0\cdot\gamma t}\eta_0(0)
\end{pmatrix}, \]
where \(\eta_k(0)\) are determined by initial conditions. This provides us with interpretation of \( \eta_k\) as two dressed states with coefficients \(\gamma\cdot \lambda_k =\gamma\cdot(1+S-(-1)^k\sqrt{S^2+\epsilon^2})\)  as their relaxation rates. We interpret the modes as \textbf{the bright and dark states}, i.e. respectively short and long living.

It is important to realize that the calculated above states \( \eta_{0,1}\) and corresponding rates \( \lambda_{0,1}\) are very general features of mathematical properties of two coupled equations (Eq.~\ref{eq:dalet}). Their consequences are seen in a wide range of problems, for example in a collisionally perturbed atomic fluorescence \cite{series1978, pendrill1978}.

\subsection{Time dependent part of Equation \ref{eq:he} in the dark and bright states basis}
In the previous section, the population evolution has been discussed under conditions when the evolution generator is independent on time. Such a situation is realized when there is just one driving field, like one \(MW\). The analysis revealed existence of two states, i.e. linear combinations of the differences between populations and stationary solution, which tend to an equilibrium with two different rates \(\gamma\cdot \lambda_k =\gamma\cdot(1+S-(-1)^k\sqrt{S^2+\epsilon^2})\) given by eigenvalues of \(\gamma\cdot\mathcal{L}_0\). These states, derived above, are defined as the difference between initial and equilibrium values, so they can be added and multiplied and thus they span a vector space in the vicinity of stationary solution. 

Here, we will present equations for a full, i.e. time-dependent evolution in the \(\eta_k\) basis. The representation of the time-dependent Liouvillian in the this basis is:

\[\mathcal{L}'_1(t)= U_R^{-1}\mathcal{L}_1(t)U_R= S \cos (\delta t)\cdot \left( 1 + \begin{pmatrix}
\sin 2\theta & -\cos 2 \theta  \\
 -\cos 2 \theta & -\sin 2 \theta
\end{pmatrix} \right)
\]
or, using the explicit form of the mixing angle \(\theta\):

\[\mathcal{L}'_1(t)=\frac{S}{\sqrt{S^2+\epsilon^2}} \cos (\delta t)\cdot\begin{pmatrix}
\sqrt{S^2+\epsilon^2}+S & -\epsilon \\
-\epsilon& \sqrt{S^2+\epsilon^2}-S
\end{pmatrix}.
\]
The above representation of \(\mathcal{L}'_1\) yields 
\begin{equation}\label{eq:waw}
\begin{split}
\dot \eta_1 &= - (1+S+\sqrt{S^2+\epsilon^2}) \eta_1 -\frac{S \cos (\delta t)}{\sqrt{S^2+\epsilon^2}}  \left( (\sqrt{S^2+\epsilon^2}+S)(\eta_1+\bar\eta_1)-\epsilon (\eta_0+\bar \eta_0)\right),\\
\dot \eta_0 &= - (1+S-\sqrt{S^2+\epsilon^2}) \eta_0 -\frac{S\cos (\delta t)}{\sqrt{S^2+\epsilon^2}} \left( -\epsilon(\eta_1+\bar\eta_1)+(\sqrt{S^2+\epsilon^2}-S)(\eta_0+\bar \eta_0)\right),
\end{split}
\end{equation}
where \(\bar \eta_1\) and \(\bar \eta_0\) are coordinates of \(\bar \eta  = U_R^{-1}\bar n\):

\[\begin{split}
\bar \eta_1 & = \frac{(1+\epsilon) \cos \theta }{\lambda_1 }n_1^0 - \frac{(1-\epsilon) \sin \theta }{\lambda_1 }n_0^0,\\
\bar \eta_0 & = \frac{(1+\epsilon) \sin \theta }{\lambda_0 }n_1^0 + \frac{(1-\epsilon) \cos \theta }{\lambda_0 }n_0^0. 
\end{split}
\]

Equations~\ref{eq:waw} demonstrate that states \(\eta_1\), \(\eta_0\) are mutually coupled by interaction with \(MWs\) or, more precisely, by beating between two \(MWs\), or \(CPO\). Interestingly, when \(S\) becomes large (\(S\gg\Gamma\)), the coupling affects the two states differently.

In the limit of very strong MWs  (\(S\rightarrow +\infty \)) one obtains \(\bar \eta_1 \rightarrow 0\), \(\bar \eta_0 \rightarrow \frac{(n_1^0+n_0^0)+\epsilon(n_1^0-n_0^0)}{\sqrt 2}\). Moreover, since \( \tan \theta \xrightarrow{S \rightarrow +\infty} 1\), the states \(\eta_1\) and \(\eta_0\) converges to the difference and sum of bare populations:

\begin{equation}
\begin{split}
 \eta_1  &\xrightarrow{S \rightarrow +\infty} \frac{n_1-n_0}{ \sqrt 2},\\
 \eta_0 &\xrightarrow{S \rightarrow +\infty} \frac{n_1+n_0}{ \sqrt 2} - \frac{(n_1^0+n_0^0)+\epsilon(n_1^0-n_0^0)}{\sqrt 2}.  
\end{split}
\end{equation}

We apply this result to the initial differential equations (Eqs. \ref{eq:dalet}) which yields:

\begin{equation}\label{eq:zain}
\begin{split}
\dot \eta_1 &= - (1+2S) \eta_1 - \epsilon \, \eta_0 - 2S\cos (\delta t)\eta_1 +\frac{1-\epsilon^2}{\sqrt{2}}(n_1^0-n_0^0),\\
\dot \eta_0 &= - \eta_0-\epsilon \, \eta_1.
\end{split}
\end{equation}

The above equations induce following observations: 
\begin{itemize}
    \item Firstly, they show that strong bichromatic field couples directly to the population difference, and only indirectly to their sum. 
    \item Secondly, in the first equation the relaxation rate in the self-coupling term (\(\lambda_1\)) grows asymptotically to \(1+ 2 S\) and dominates the cross-coupling term \(\epsilon \, \eta_0\).  Consequently, the difference of the initial populations exhibits a periodic CPO dynamics driven by oscillations proportional to \(2 S \cos \delta t\).
    \item Further, in the second equation the self-coupling constant converges to \(1\) which is the limit of \(\lambda_0\) for \(S\) approaching infinity.
\end{itemize}
Which triggers several conclusions (considered in the three regimes of the MWs detuning): 
\begin{itemize}
    \item \textbf{For \(\delta\simeq \gamma\)} both states exhibit a periodic CPO dynamics.  The amplitude of \(\eta_1\) is proportional to the source term \(\frac{1-\epsilon^2}{\sqrt{2}}(n_1^0-n_0^0)\). State \(\eta_0\) is indirectly driven by force oscillating with the MWs detuning via the cross term \(\epsilon \, \eta_1\). Still, its oscillations are insignificant compared to the amplitude of \(\eta_1\) and restricted to the MWs detuning smaller than \(\gamma\). This results in the composite resonance and justifies why the width of the resonance component corresponding to \(\eta_0\) is limited from above by \(\gamma\) and its amplitude decreases monotonically with \(S\).
    \item \textbf{For \(\Gamma \gg \delta \gg\gamma\)} the cross-coupling term \(\epsilon \,\eta_1\) in the second equation oscillates with a frequency much higher than the relaxation rate determined by stabilization of the self-coupling constant \(\lambda_0\) for strong MWs, hence it can be neglected. Since the dynamics of  state \(\eta_0\) is not driven by any source term its amplitude for \( \delta \gg\gamma\) vanishes exponentially in time. The sum of populations becomes constant and the system effectively closes. Consequently, the cross term \(\epsilon \, \eta_0\) in the first of Equations \ref{eq:zain} becomes zero and states \(\eta_0\) and \(\eta_1\) can be regarded as decoupled for strong MWs.
    \item \textbf{For \(\delta\simeq \Gamma\)} the driving term \(2S\cos \delta t\) oscillates much faster than \(1+2S\). Consequently, it has negligible effect on the dynamics of state \(\eta_1\) which is governed solely by the source term \(\frac{1-\epsilon^2}{\sqrt{2}}(n_1^0-n_0^0)\). Thus, for a large detuning of the MWs the amplitude of \(\eta_1\) becomes stationary at \(\langle \eta_1\rangle \sim \frac{n_1^0-n_0^0}{1+2S}\) and one recovers the regular ODMR resonance shape.
\end{itemize}

Importantly, Eqs.~\ref{eq:zain} provide mathematical evidence and interpretation of the observations presented in the main paper - the effect of the power stabilization of the narrow resonance. This also justifies our interpretation of the uncoupled state as the bright one and the coupled state as the dark one.

The described effect is to some extent analogous to the effect of CPT. An important difference, though, is that here we consider combinations of populations, rather than wave functions or coherences in the familiar interpretation of the CPT effect \cite{gray1978, arimondo1996}.

\section{First harmonic approximation, resonance widths}

By substitution of  \( n_i(t)= \alpha_i +\beta_i \cos(\delta t +\phi_i) \,\, \forall i=0,1 \) into Eqs. \ref{eq:dalet}, ignoring the time dependence of all expressions except of constants and quantities oscillating as \(e^{\pm i \delta t}\), we arrive at three pairs of equations for  \(\alpha_i, \, \beta_i,\, \phi_i\). As stated in the main article, we take the fluorescence intensity as proportional to the time-averaged population difference \( I \sim \langle n_1-n_0 \rangle = \alpha_1-\alpha_0\) and present the fluorescence as composed of several contributions, two of them exhibiting resonance dependence on \(\delta\):
\begin{equation}
I \sim \Delta n^0 \cdot \frac{1}{1+\frac{2S}{1-\epsilon^2}}\cdot \left(1-A_0 \frac{w_0^2}{\delta^2+w_0^2}-A_1 \frac{w_1^2}{\delta^2+w_1^2}\right).
\end{equation}
By making substitutions
\[\begin{split}
\tau & = \frac{\lambda_1^2+\lambda_0^2}{2}-\frac{S^2(1+2S)}{1+2S-\epsilon^2},\\
\mu & = \frac{\lambda_1^2-\lambda_0^2}{2}-\frac{S^2(1+2S)}{1+2S-\epsilon^2},\\
\nu & = \frac{4S^2(1+S)\epsilon^2}{\left(\sqrt{S^2+\epsilon^2}+S\right)\left(\sqrt{S^2+\epsilon^2}+1+S\right)}.
\end{split}\]
the widths \(w_0\) and \(w_1\) and amplitudes \(A_0\) and \(A_1\) of the two last resonance contributions can be cast in simple forms: 
\begin{equation}\begin{split}
\Big(\frac{w_0}\gamma\Big)^2  &= \tau - \sqrt{\mu^2+\nu},\\
\Big(\frac{w_1}\gamma\Big)^2  &= \tau + \sqrt{\mu^2+\nu},\\
A_0  & =\frac{S^2}{\sqrt{\mu^2+\nu}}\cdot \Big( \Big(\frac{\gamma}{w_0}\Big)^2-\frac{1+2S}{1+2S-\epsilon^2}\Big),\\
A_1 & =\frac{-S^2}{\sqrt{\mu^2+\nu}}\cdot \Big(\Big(\frac{\gamma}{w_1}\Big)^2-\frac{1+2S}{1+2S-\epsilon^2}\Big).
\end{split}
\label{eq:widthsamps}
\end{equation}

Eqs.~\ref{eq:widthsamps} can be further simplified by considering a strong field limit (\(S\gg\)1) and making the approximation \(\frac{\nu}{2\mu}\simeq 0\). Then the widths can be expressed with very simple formulae:
\begin{equation}
\begin{split}
w_0&\simeq \gamma \cdot \lambda_0\\
w_1&\simeq\gamma \cdot \sqrt{\lambda_1^2-2\frac{S^2(1+2S)}{1+2S-\epsilon^2}}\simeq \gamma\cdot \sqrt{\lambda_1^2-2S^2}.
\end{split}\label{eq:widths}
\end{equation}

Equations~\ref{eq:widths} represent the power dependence of linewidths of the composite resonance discussed in Figs. 3 and 5 of the main paper.

%